\documentclass{aa}  

\newcommand{\UU}{\mathbf{u}}
\newcommand{\BB}{\mathbf{B}}

\newcommand{\Rmax}{R_\mathrm{max}}
\newcommand{\NH}{N_\mathrm{H}}
\newcommand{\Sd}{\mathcal{S}}

\newcommand{\kmpss}{\mathrm{km} \, \mathrm{s}^{-2}}
\newcommand{\parsec}{\mathrm{pc}}

\newcommand{\za}{z_\mathrm{a}}

\newcommand{\jnu}{j_\nu}
\newcommand{\jnuz}{j_{\nu,0}}

\newcommand{\cbin}{C_\mathrm{bin}}
\newcommand{\cbini}{C_\mathrm{bin,i}}

\newcommand{\PIX}{{PlanckXIX}}
\newcommand{\PXX}{{PlanckXX}}

\usepackage{color}
\usepackage{subcaption}
\usepackage{gensymb}
\usepackage{amsmath,amssymb}
\usepackage{bm}
\usepackage{mathtools}
\usepackage[bookmarks=false]{hyperref}
\def\blue{\textcolor{black}}
\definecolor{purple}{rgb}{0.5,0,.5}

\usepackage{graphicx}
\usepackage{txfonts}
\begin{document}

   \title{The supernova-regulated ISM. IV. A comparison of simulated polarization with Planck observations}
   \titlerunning{The SN-regulated ISM. IV. Simulated polarization}

   \author{M.S. V\"ais\"al\"a
          \inst{1,2} \fnmsep\thanks{The author's affiliation since January 2018: Academia Sinica, Institute of Astronomy and Astrophysics, Taipei, Taiwan}
          \and
          F.A. Gent\inst{2} \and 
          M. Juvela\inst{1} \and
          M. J. K\"apyl\"a \inst{3,2} \fnmsep\thanks{This work belongs to the Max Planck Princeton Centre for Plasma Physics framework}
          }

   \institute{Department of Physics, Gustaf H\"allstr\"omin katu 2a, PO Box 64, 00014 University of Helsinki, Finland\\ 
              \email{miikkavaisala@gmail.com}
         \and
             ReSoLVE Centre of Excellence, Department of Computer Science, 
	      Aalto University, PO Box 15400, FI-00076 Aalto, Finland
             \and
             Max Planck Institute for Solar System Research,
              Justus-von-Liebig-Weg 3, D-37077 G\"ottingen, Germany \\
             }

   \date{Received / Accepted}

% \abstract{}{}{}{}{} 
% 5 {} token are mandatory
 
  \abstract
   {The efforts for comparing dust polarization measurements with synthetic observations
    from MHD models have previously concentrated on the scale of molecular clouds. }
   {Here we extend the model comparisons to kiloparsec scales, taking into account
   hot shocked gas generated by supernovae, and a non-uniform
   dynamo-generated magnetic field at both large and small scales 
   down to 4\,pc spatial resolution.}
   {Radiative transfer calculations are used to model dust emission and polarization
   on the top of MHD simulations. We compute synthetic maps of column density
   $\NH$, polarization fraction $p$,
   and polarization angle dispersion $\Sd$, and study their dependencies on
   the important properties of the MHD simulations. 
   These include the large-scale magnetic field and its orientation, the small-scale magnetic field,
   and supernova-driven shocks.}
   {Similar filament-like structures of $\Sd$ 
   as seen in the Planck all-sky maps are visible in our synthetic results,
   although the smallest scale structures are absent from our maps.
   Supernova-driven shock fronts and $\Sd$ do not show significant correlation.
   Instead, $\Sd$ can clearly be attributed to the distribution of the
   small-scale magnetic field.
   We also find that the large-scale magnetic field influences the polarization
   properties, such
   that, for a given strength of magnetic fluctuation, a strong plane-of-the-sky
   mean field weakens the observed $\Sd$, while strengthening $p$. 
   The anticorrelation of $p$ and $\Sd$,
   and decreasing $p$ as a function of $\NH$ are consistent across all synthetic observations.
   The magnetic fluctuations follow an exponential
   distribution, 
   rather than Gaussian, characteristic of flows with intermittent 
   repetitive shocks.}
   {The observed polarization properties and column densities are sensitive to
   the line of sight distance over which the emission is integrated.
   Studying synthetic maps as
   the function of maximum integration length will further help with the interpretation
   of observations.
   The effects of the large-scale magnetic field orientation on the polarization
     properties are difficult to be quantified from observations solely, but MHD models
   might turn out to be useful for separating the effect of the large-scale mean field.
   }

               
   \maketitle
%
%-----------------------------------------------------------------------------
\section{Introduction}
%-----------------------------------------------------------------------------

Magnetic fields are dynamically important constituents of galaxies, playing a
major role, e.g., in the star-formation process and controlling the density and
propagation of cosmic rays
\citep[see, e.g.,][and references therein]{beck2016}.
Observing them, however, is non-trivial, as indirect observations are required,
based primarily on dust polarization, Zeeman 
effect, synchrotron radiation, its polarization, and
Faraday rotation of the polarization angle
\citep[referred to as rotation measure, hereafter RM -- see e.g.][and references therein]{KleinFletcher2015}.
Because all such methods have strong limitations, interpretation of the data
is difficult, especially for the Milky Way, inside of which we reside. 
This is where radiative transfer simulations combined with numerical models may
become useful, bridging the differences between
physical models and indirect observations
\citep[e.g.][hereafter \PXX]{ostriker2001,FaGo2008,pelkonen2009,Planck2015XX}.

Planck is a space mission that mapped the anisotropies of the cosmic
microwave background \citep[CMB, see e.g.][]{Planck2011I,Planck2015Ires}.
It also has the capability to measure thermal emission and its polarization
from dust grains in all bands up to 353\,GHz.
Particularly, with the 
High Frequency Instrument in the frequency 
range 100--857\,GHz \citep[HFI, see e.g.][]{lamarre2010}, the foreground dust can be studied. 
Polarized dust emission and its spatial variations have been mapped with high
resolution and sensitivity in a series of papers.
\citet[][\PIX\ hereafter]{Planck2015XIX} study the all-sky dust emission at
353\,GHz, where polarized emission is most significant, and \PXX\ compute
the statistics of polarization fractions and angles outside the galactic plane.
In \citet{Planck2015XXI} thermal dust emission is compared with optical
starlight polarization.
\citet{Planck2015XXII} presents a study of the variation of dust emission as a
function of frequency in the range 70--353\,GHz.
The results relating to the polarized thermal dust emissions are summarised 
in \citet[][Section 11.2]{Planck2015Ires}.
For this paper, the most relevant study in this series is the all-sky study
of \PIX\ (and subsequent updates reported in \citet{Planck2016X}),
as our modelling efforts concentrate on
kiloparsec (kpc) scale magnetohydrodynamic (MHD) models including all the
three phases (cold, warm, hot) of the interstellar medium (ISM), regulated by
supernova (SN) activity and subject to large- and small-scale dynamo
instabilities.
Here cold would correspond to cold neutral medium (CNM) with typical temperatures of
100\,K and number densities of $\sim$100\,cm$^{-3}$, 
warm to 10$^4$\,K and 0.1\,cm$^{-3}$, and hot to 10$^6$\,K and 0.001\,cm$^{-3}$.
The major findings of \PIX\ 
 include the discovery of
anti-correlation between the polarization fraction, $p$, and polarization
angle dispersion, $\Sd$, and the decrease in the maximum polarization fraction,
$p_{\rm max}$, as column density increases.
\blue{These major findings generally hold in zoomed-in regions near
molecular cloud complexes (\PXX).}

The efforts for comparing polarization measurements with synthetic observations
from MHD models has concentrated on the scale of molecular clouds.
For such comparisons, the relevant MHD models normally include the cold and
warm phases of the ISM, describe the magnetic field as a uniform background
field, and may include artificial flows to enhance the formation of shocks
\citep[e.g.][]{ostriker2001,PGDJNR01,BCLK07,FaGo2008,hennebelle2008,SHMMNF13,
Planck2015XX,CKL16}.
Appropriately the Planck data provide high enough sensitivity and
resolution for studies at this length scale.
In general, a satisfactory agreement between the synthetic and observed dust
polarization properties has been found, with
the anisotropic and turbulent character
of the magnetic field having been identified as the most decisive factor, particularly \PXX\ have demonstrated the connection of $p$ and $\Sd$ with turbulent magnetic fluctuations.

The large-scale dynamics of the ISM in the star-forming parts of spiral 
galaxies can be described with a three-phase medium regulated by stellar energy input \citep[][]{MO77}. 
By far the dominating energy source to power turbulence at the 100\,pc scale 
\citep[see][]{Abbott82} originates from
supernova explosions (SNe). SNe bring significant input of
thermal and kinetic energy to the ISM. In the solar neighbourhood \citet{TLS94}
estimate that SNe inject approximately
$3\times10^{52}$\,erg\,kpc$^{-2}$\,Myr$^{-1}$ thermal energy, which is 
dissipated mainly as heat into the ISM, but with
some 10\% converted into kinetic energy \citep{Chevalier77,Lozinskaya92}. In addition to driving
expanding shock fronts that interact with each other, the SNe generate bubbles
of hot gas near the galactic disk,
which as well as the cold ISM and molecular clouds are embedded within the 
diffuse warm component.
The filling factor of the hot gas is small near the galactic midplane but
approaches unity in the halo \citep[][]{ferriere2001}. 

In addition to the re-structuring and mixing of the ISM, SN-forcing powers the
galactic dynamo in the rotating anisotropic galactic disk.
Anisotropic turbulence and a non-uniform rotation profile combine to
provide the ingredients for the large-scale dynamo instability, leading to the
generation and maintenance of magnetic fields dynamically significant on the
galactic scale. 
Along with the mean magnetic field, a strong fluctuating field is also 
generated.
The dynamo processes are intrinsically connected to the three-phase structure of
the ISM, so that both the large- and small-scale filling factor and topology
are different in various phases and locations of the galactic disk.
Recent numerical MHD models have
attained sufficient realism to model these processes 
self-consistently \citep{gent2013II,HI14,KO15,BGE15,EGSFB17a,HSSFG17}.
These developments enable us to study the influence of the three-phase medium,
SN shock fronts, and dynamo-generated magnetic fields on the observable
properties of dust polarization at large scales.

In this work we study the influence of all the aforementioned physical effects 
on the polarization properties of the galactic ISM. Apart from \PXX\ and
\citet{PlanckXXXV16}, our approach is different
from most of the previous modelling efforts related to Planck observations in that we are not
building models to specifically explain the observations, but synthesise a set
of independently-built turbulence models to test the relevance of the physical effects
they contain to explain real observations.
Our strategy is also to be contrasted with
studies that build statistical or phenomenological models to the observed
all-sky polarisation properties \citep[e.g.][]{Planck2016XXXII,Planck2016XLII,Planck2016XLIV}.

To be able to resolve, on the one hand,
the SN-generated turbulence and, on the other, to allow for self-consistent dynamo
action on kiloparsec scales, the simulation setup is limited by these
requirements, and we are therefore not free to arbitrarily choose the resolution optimally suited to the observations.
In this study, in particular, we are limited to 4\,pc grid resolution
with the kiloparsec box size chosen.

The chosen resolution in practice means that we can model the cold cloud component only up to
certain densities, therefore preventing them from becoming gravitationally bound.
The main regulatory mechanism creating the cold and warm phases is, however, properly included, namely the
thermally unstable cooling function. Therefore,
we consider having a realistic description of the three-phase medium. At the
box scale, the dimensions are yet too small for the large-scale effects such
as spiral arms or central bulge to be realistically included.
We note that in contrast \citet{Planck2016XLII}, who build a phenomenological whole-sky model of the Milky Way, do model the spiral
arms, but they do not include a physically self-consistent model of turbulence.
Therefore, we restrict ourselves to consider only large-scale effects arising from
rotation and its non-uniformities (differential rotation), both of which are
needed to enable and sustain dynamo action in the system.

The paper is organized as follows. In Sect.\,\ref{sec:methods} we describe
the tools and methods used in this study. In Sect.\,\ref{sec:results} 
we present the simulated polarization results and compare them to the
observations of \PIX. 
In Sect.\,\ref{sec:analysis} we consider how the shock and magnetic field 
affect the interpretation of observations.
In Sect.\,\ref{sec:conclusions}, concluding the
paper, we discuss the implications of our results and potential for further
studies.  

%-----------------------------------------------------------------------------
   \begin{figure*}
   \centering
      \begin{subfigure}[t]{0.32\textwidth}
    \includegraphics[
                     trim=1.75cm 0.2cm 1.75cm 1.75cm,
                     clip=true,
                     width = \hsize]{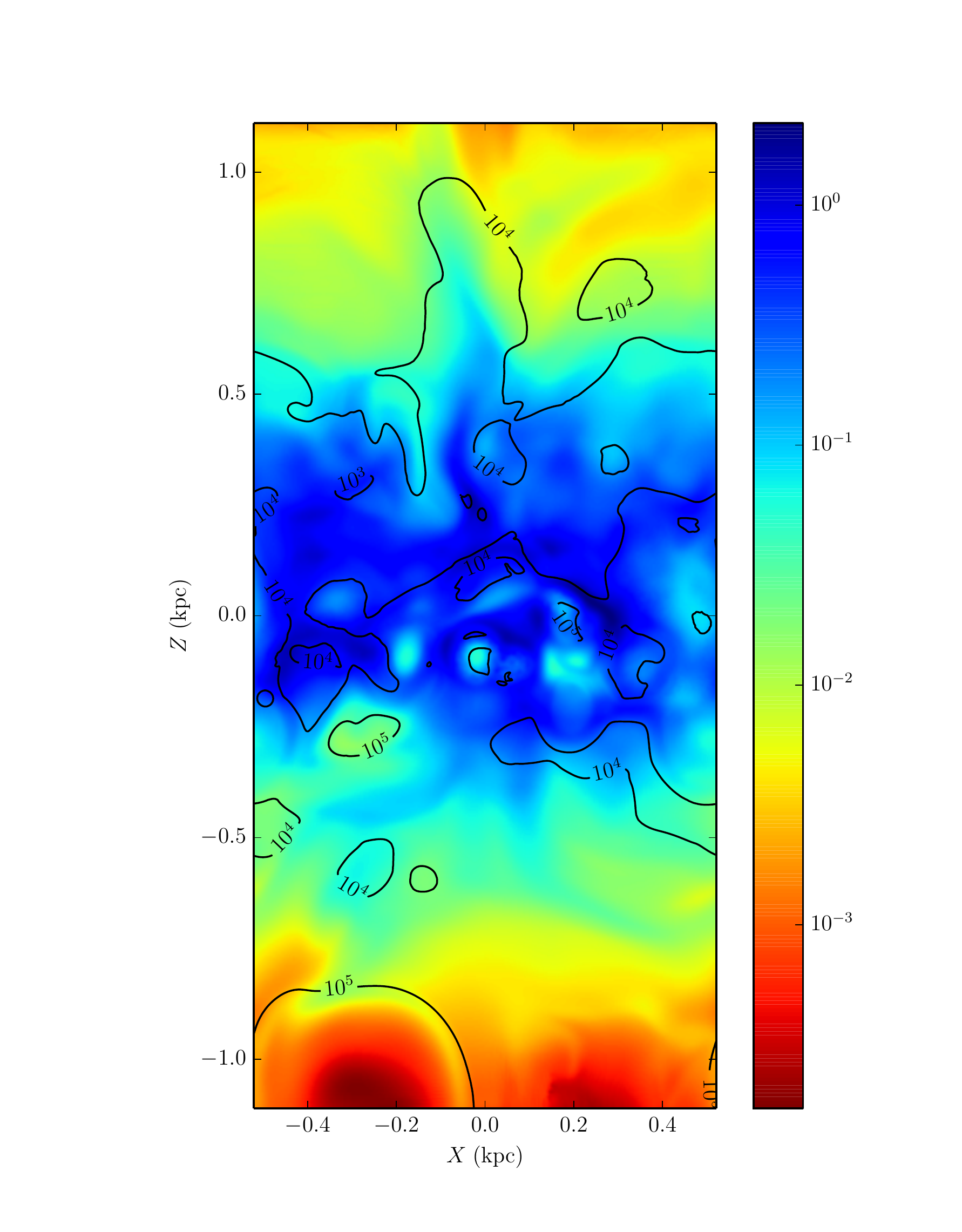}
      \end{subfigure}
      \begin{subfigure}[t]{0.64\textwidth}
    \includegraphics[
                     trim=0.5cm 0.2cm 0.5cm 0.75cm,
                     clip=true,
                     width = \hsize
                    ]{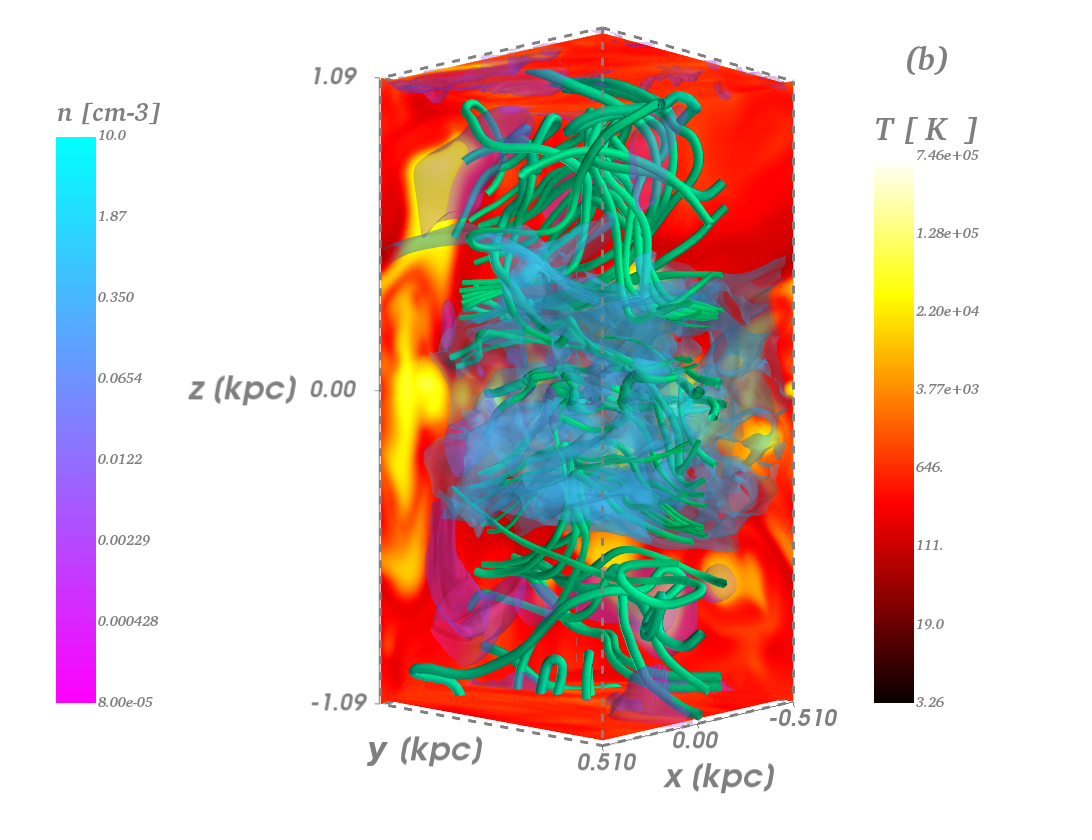}
      \end{subfigure}
      \caption{
      {\it Left:} Representative slice in the $xz$-plane of gas number density $n\,[1/\mathrm{cm}^{-3}]$
      from a single snapshot overplotted with contours of temperature $T$\,[K],
      illustrates the multi-phase ISM.
      {\it Right:} Snapshot of the model ISM, with temperature slices in the 
      background and isosurfaces of the density in the foreground, with 
      fieldlines of the magnetic field over plotted. 
         }\label{fig:rhoT}
    \label{fig:Bfield}\label{fig:velocity_field}
  \end{figure*}
%-----------------------------------------------------------------------------

%-----------------------------------------------------------------------------
   \begin{figure*}
   \centering
      \begin{subfigure}[t]{0.48\textwidth}
         \includegraphics[trim=0.0cm 0.0cm 0.75cm 0.0cm,clip=true,height = \hsize]{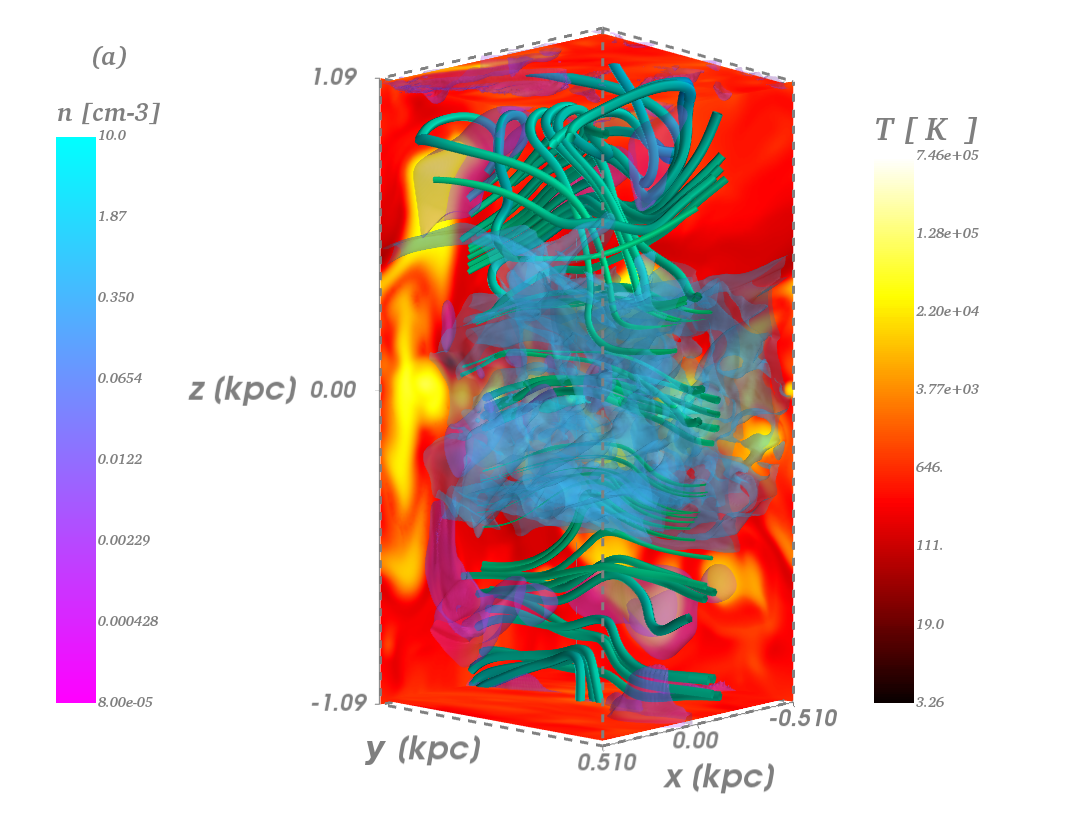}
      \end{subfigure}%\hspace{-0.2cm}
      \begin{subfigure}[t]{0.48\textwidth}
         \includegraphics[trim=10.5cm 0.0cm 0.0cm 0.0cm,clip=true,height = \hsize]{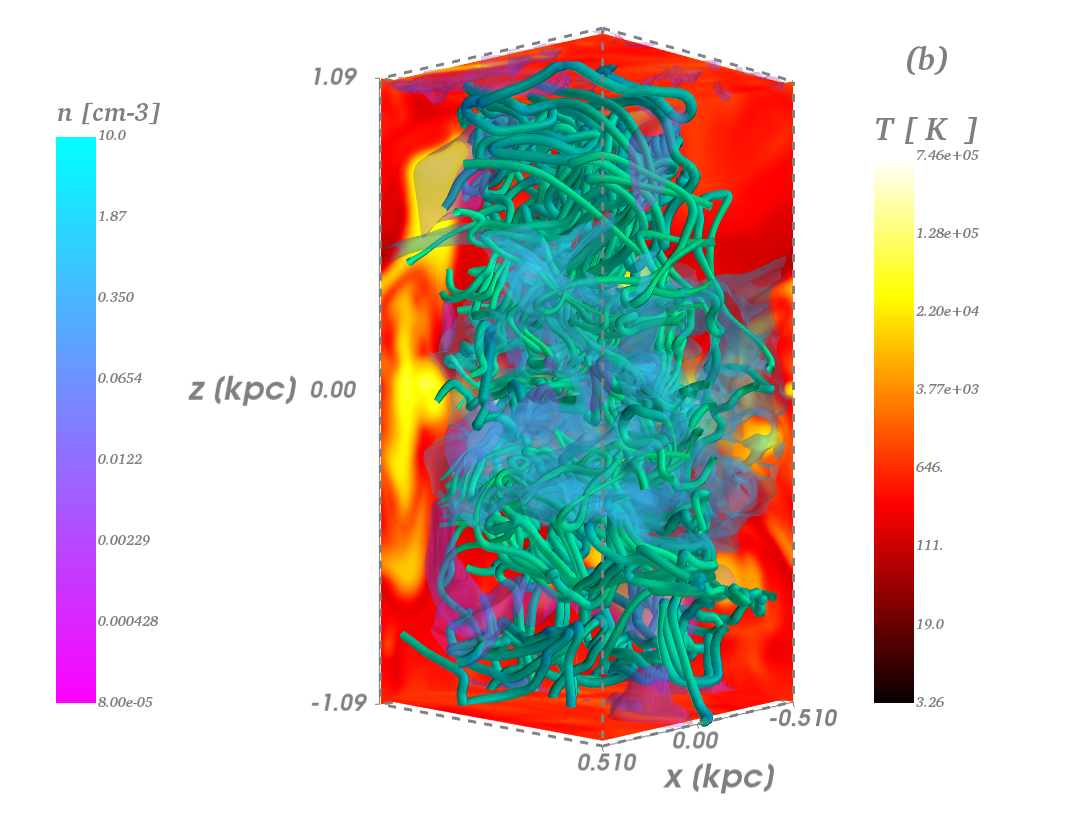}
      \end{subfigure}
      \caption{
      Snapshot of the model ISM, with temperature slices in the 
      background and isosurfaces of the density in the foreground, 
      as in Fig\,\ref{fig:Bfield}, but with the total magnetic 
      $\mathbf{B}$ fieldlines replaced by $\mathbf{\bar{B}}$ fieldlines ({\it left})
      and $\mathbf{b}$ fieldlines ({\it right}). 
      For more visualisation of the MHD data, including video representation, 
      visit \href{http://fagent.wikidot.com/astro}{http://fagent.wikidot.com/astro}
               }
         \label{fig:3Dplots}
   \end{figure*}
%-----------------------------------------------------------------------------

%--------------------------------------------------------------------
\section{Methods}\label{sec:methods}
%-----------------------------------------------------------------------------

%-----------------------------------------------------------------------------
\subsection{Numerical MHD simulations}\label{sect:mhdsim}

In \PXX, polarization statistics are compared to MHD
simulations, which include cold and warm phases of the ISM. 
These employ adaptive mesh refinement in a computational cube 50 pc across
\citep{hennebelle2008}, from which an $18^3$\,pc$^3$ subset is 
selected for analysis.
Here, we add comparisons to MHD simulations of the ISM, in
which the turbulence is driven by SNe 
\citep[][Chapters 8 \& 9]{gent2013II,Gent2012}.
In this model, the cold and warm phases are produced, as in the two-phase
models, through regulation by thermally unstable radiative cooling,
but with the addition of a hot phase generated by SN heating. 

To capture all the relevant dynamics of the three-phase model, the simulation
domain size has to be increased.
The grid is $256 \times 256 \times 560$ and spans horizontally 1.024\,kpc and
vertically $\pm1.12$\,kpc about the galactic midplane.
The supernova supersonic forcing naturally generates a highly shocked 
turbulent flow, so no artificial forcing is applied.
Moreover, the interaction of rotation and anisotropic turbulence with the 
galactic shear flow induces a natural magnetic field through dynamo action.
To model dynamo action, we solve non-ideal MHD, including 
viscous, thermal and magnetic diffusivities.
With temperatures spanning 7--8 orders of magnitude, and high Mach numbers, it 
is not possible to apply the physically motivated values for diffusivity. 
To resolve the flows in the hot gas, while obtaining optimal small scale 
structure to the turbulence in the cold and warm ISM, we set the viscosity
proportional to the sound speed (or $\sqrt T$), which may be significant for
analysis in Sect.\,\ref{sec:NH}. This is not chosen to model the actual
   turbulent diffusivities,
  as little is known about their dependence on the key physical quantities in the ISM.
  The analytical estimates obtained in the first-order smoothing approximation
  framework \citep{SKR66}, however, imply that the magnitude of turbulent
  diffusion is orders of magnitude larger than the Spitzer molecular values.
  The motivation for the used viscosity scheme is to resolve much finer structures in the cool and warm ISM, 
than would be possible with constant diffusivity.

Note, that the resolution of $4$\,pc along each side in these simulations
without magnetic fields \citep[][Table\,5.1]{Gent2012} yield a maximum gas
number density for the ISM of about $ 100\, {\rm cm}^{-3}$ and the fractional
volume of the cold gas is 0.4\%.
The fractional volume of the warm gas is 60\%, with hot gas about 28\% near the
SN active midplane and increasing to 41.5\% elsewhere. 
With the magnetic field included \citep{EGSFB17b} the model fractional volume
of the warm gas increases to about 80\%, the hot gas being pushed further into
the halo, and the cold gas is confined to an even smaller fractional volume. 
A snapshot of the thermodynamic profile of the model ISM is displayed in
Fig.\,\ref{fig:rhoT} (left), wherefrom the distribution of the different gas phases
is evident.
Observed densities and those arising from the MHD simulations of 
\citet{hennebelle2008} extend to much higher densities and increased 
fractional volume of cold ISM.
The characteristic properties of the MHD simulation data are listed in 
Table\,\ref{tbl:snapshots}.
Temperatures, velocities and magnetic field strengths are better representatives
of the observed ISM, but smallest scales of their fluctuations are limited by
the grid resolution.
The saturation of the magnetic field has the effect of restricting the flow and
increasing the homogeneity of the ISM, so that the maximum density reduces 
to about 10\,cm$^{-3}$, which must be 
taken into account when making comparisons with the Planck observations and the
earlier MHD molecular cloud model.
Mach number in the simulations reaches as high as 25.

%-----------------------------------------------------------------------------
\begin{table}
\centering
\begin{tabular}{|c|c|rcl|rcl|}
\hline 
         &                       &            &\!\!\!\!\!\!max\!\!\!\!\!\!&             &               &\!\!\!\!\!\!min\!\!\!\!\!\!&                 \\
\hline                                                                                                                                   
$n$      &$[\mathrm{cm}^{-3}]$   &           5&\!\!\!\!\!\!-- \!\!\!\!\!\!&8.5          &$7\cdot10^{-6}$&\!\!\!\!\!\!-- \!\!\!\!\!\!&$2.7\cdot10^{-4}$\\
$T$      &          [K]          &$7\cdot10^7$&\!\!\!\!\!\!-- \!\!\!\!\!\!&$3\cdot10^8$&             135&\!\!\!\!\!\!-- \!\!\!\!\!\!&311              \\
         &&&&&&& \\                                                                                                                                   
$B$      &$[\mu\mathrm{G}]$      &6           &\!\!\!\!\!\!-- \!\!\!\!\!\!& 10          &               &\!\!\!\!\!\!0  \!\!\!\!\!\!&                 \\
$\bar{B}$&$[\mu\mathrm{G}]$      &4.6         &\!\!\!\!\!\!-- \!\!\!\!\!\!&  5.5        &               &\!\!\!\!\!\!0  \!\!\!\!\!\!&                 \\
$b$      &$[\mu\mathrm{G}]$      &4           &\!\!\!\!\!\!-- \!\!\!\!\!\!&  7          &               &\!\!\!\!\!\!0  \!\!\!\!\!\!&                 \\
$u$      &$[\mathrm{km\,s}^{-1}]$&147         &\!\!\!\!\!\!-- \!\!\!\!\!\!&  549        &               &\!\!\!\!\!\!0  \!\!\!\!\!\!&                 \\
\hline
\end{tabular}
\caption{For the 12 snapshots included in the analysis, at intervals of 
  25\,Myr, the ranges of gas number density $n$, temperature $T$,
  total $B$, mean $\bar{B}$, fluctuating $b$ magnetic field and speed $u$.} 
\label{tbl:snapshots}
\end{table}

Both large- and small-scale dynamo instabilities are present in the system.
A 3D rendition of the magnetic fieldlines embedded in this atmosphere is
illustrated in Fig.\,\ref{fig:Bfield} (right).
As reported by \citet{EGSFB17a}, the strength of the generated mean magnetic
field at the midplane in the MHD model is in good agreement with the global
observational estimates of 1.6$\mu$G by \citet{RK89}, using pulsar RM.
The random field in the model, however, is much weaker than observed, being only
20--50\% of the 5$\mu$G estimate of \citet{RK89} or 5.5$\mu$G of 
\citet{Haverkorn15}, combining RM with thermal electron density measures. 
These estimates are supported by the reviews of galactic magnetic fields 
\citep{Beck96, beck2016}.
Some preliminary results applying synthetic RM measurements to the MHD
model considered here are reported in \citet{HSSFG17}, and further such 
application shall be considered in future work. 
The model and the characteristics of the multi-phase structure of the simulated
ISM are described in detail in \citet{gent2013I}, and a summary is
 included in Appendix\,\ref{app:model}. 
The coherence and fluctuations of the magnetic field are important to the 
polarization measurements, so it is useful
to decompose the field into the mean field $\mathbf{\bar{B}}$ and fluctuations
$\mathbf{b}$, where 
\begin{equation}\label{eq:meanB} 
\mathbf{B}=\mathbf{\bar{B}}+\mathbf{b}.
\end{equation}
In this model, the entire field is subject to stirring due to the action of the
many SN remnants, so the convention of separation into large-scale and turbulent
components needs more careful consideration.
We separate $\mathbf{B}$ by volume averaging with a Gaussian kernel 
\citep[At $l = 50$\,pc scale, see][]{gent2013II}.
In Fig.\,\ref{fig:3Dplots}
(left)
$\mathbf{\bar{B}}$ fieldlines and 
(right)
$\mathbf{b}$ fieldlines are plotted 
over density isosurfaces on background slices of temperature.
With this treatment the `mean field', which is what we shall refer to as
the \emph{large-scale} has strength and orientation which varies with space and
time,
but is \emph{coherent}, i.e. similarly oriented, over scales above about 100\,pc. 
The \emph{small-scale} field is derived by subtracting the large-scale field
from the total field for any given snapshot.
So the large-scale field still exhibits spatial structure, which would influence
polarization fractions (see Sect.\,\ref{sec:rtc}).
The random field is clearly incoherent and would contribute to 
depolarization.
Some indicative values are listed in Table\,\ref{tbl:snapshots} for the
ranges across the MHD snapshots of gas number density, temperature,
magnetic field and speed.

The gas density $\rho$ from each snapshot is used with the radiative transfer
code SOC, as described in Sect.\,\ref{sec:rtc},
to model the dust temperature distribution, and the 
magnetic field $\BB$ is then employed when simulating the dust polarization
observations. 
From the velocity field we can compute directly its divergence, 
$\nabla\cdot\UU$, which is also used in the analysis of the polarization to
determine the influence of the shock structure of the ISM on the observations.

The simulation used for this analysis applies parameters for gas density, 
stellar and dark halo gravitation, galactic rotation, and 
SN rates and distribution matching estimates for the Solar neighbourhood. 
The magnetic field is amplified for a period exceeding a Gyr by dynamo
action from a seed field of a few nG, which then saturates with an
average field strength of a few $\mu$G.
The strength of the mean field portion is consistent with observational 
estimates, while the random field strength is 2--5 times weaker than 
estimated.
This has some influence on the dispersion of polarization angles in the simulated
observations, discussed in Sect.\,\ref{sec:analysis}.
We use 12 snapshots from the saturated statistical steady stage of the model,
each separated by 25\,Myr, commencing at 1.4\,Gyr.
For simulated maps presented in this paper, we use the snapshot at 1.7\,Gyr. 
The system is in a statistical steady state, so individual snapshot
characteristics are representative, but strong temporal and spatial differences
are also present.
For some of the analysis we consider ensemble averages of the snapshots to
identify the most persistent structures. 

%-----------------------------------------------------------------------------
\subsection{Stellar radiation field}\label{sec:isrfmodel}
%-----------------------------------------------------------------------------

Dust in the ISM is illuminated by the stellar radiation field.
We invert the measurements of the average stellar radiation field in the solar
neighbourhood from \citet{mathis1983} into a distribution of
radiation sources, to model the stellar radiation for our radiative
transfer model.
We then model this radiation with a horizontal density profile
\citep[see][excluding the dark matter component]{gent2013I}, and
emissivity $\jnu$ reflecting the vertical distribution of stars as
\begin{equation}
   \jnu(z) = \jnuz \exp\bigg[a_1 \bigg(\za - \sqrt{\za^2+z^2}\, \bigg) \bigg].
   \label{brightness_dist}
\end{equation}
Here $a_1 = 4.4 \cdot 10^{-14}$ $\kmpss $, $\za = 200.0$ $\parsec$ and 
$\jnu$ is the emissivity over the spectrum of frequencies $\nu$.
The normalization coefficients, $\jnuz$, are determined to return the expected
total intensity $J_\nu$ from 
\begin{equation}
    J_\nu = \frac{V}{4 \pi N} \sum_i^{i=N} \frac{\jnu(z_i, \jnuz = 1)}{4\pi d} e^{-\tau_i}
\end{equation}
where $d = \sqrt{z^2 + r^2}$, $V = 2 \pi Z_0 R_0^2$ and $\tau$ is the
optical depth along the line of sight (LOS).
The distribution is generated using Monte-Carlo method, and it is scaled to
match $J_{\nu\mathrm{, Mathis}}$ from \citet{mathis1983}.
Using this inversion, we obtain an approximate $z$-dependent radiation field,
which produces reasonable dust temperatures with our radiative transfer
simulations.

%-----------------------------------------------------------------------------
\subsection{Radiative transfer calculations}\label{sec:rtc}
%-----------------------------------------------------------------------------

To calculate the dust emission, we use the program SOC, which 
is a new Monte Carlo code for continuum radiative transfer
against the CRT program \citep{Juvela2005} and also against several
other codes participating in the
TRUST\footnote{http://ipag.osug.fr/RT13/RTTRUST/} benchmark project on
3D continuum radiative transfer codes \blue{\citep{Gordon2017}}. The
density distribution of the models can be defined with regular or
modified Cartesian grids or hierarchical octree grids. In this paper,
all calculations employ regular Cartesian grids. In addition to the
density field, the program needs a description of the dust grains
(i.e. absorption and scattering properties) and of the radiation
sources. The program employs a fixed frequency grid to simulate the
radiation transport at discrete frequencies, one frequency at a time.
The information of the absorbed energy is used to solve the grain
temperatures for each cell of the model. The dust model could also
include small stochastically heated grains. However, to predict dust
emission at submillimetre wavelengths, the calculations are here
limited to large grains that are assumed to remain at a constant
temperature, in an equilibrium with the local radiation field. Once
the dust temperatures have been solved, synthetic images of dust
emission can be calculated towards selected directions or, as in the
case of the present study, over the whole sky as seen by an observer
located inside the model volume. 

We assume a constant gas-to-dust ratio, where we follow the dust model
BARE-GR-S of \citet{Zubko2004}, which has been created to match the
observations of typical Milky Way dust with the extinction factor $R_v=3.1$. For simplicity, we keep
the properties of the dust similar throughout the whole computational domain. 
Dust properties may vary between different ISM phases, but the investigation of these 
effects is beyond the scope of this initial study.
 
During the simulation of the internal radiation field, the radiation
transport is calculated without taking the polarization into account.
However, SOC includes tools to produce synthetic polarization maps.
The local grain alignment efficiency could be calculated following
the predictions of the radiative torques theory
(\citet{DraineWeingartner1996}; \citet{pelkonen2009}; \PXX). Because our study
concentrates on emission from low-density medium, this step
is omitted and we essentially assume a constant dust grain alignment efficiency.
The polarization reduction factor $p_0$ is simply set to a
constant value that results in a maximum polarization fraction that is
consistent with observations. 
Maps are calculated
separately for the Stokes $I$, $Q$, and $U$ components, taking into account
the local total emission, the local magnetic field direction, and the
value of $p_0$. These data are then finally converted to maps of
polarization fraction and polarization angle dispersion.
A representative set of synthetic Stokes $I$, $Q$, and $U$ maps
is presented
in Appendix\,\ref{sec:iqu}.

To calculate the polarization within a single cell, we apply the
following method. We use the Planck/HEALPix
convention for the polarization angle $\psi$
\citep{Planck2015XIX,Planck2015XXI}.
Within a cell we normalize the direction of the magnetic field,
\begin{equation}
\mathbf{\hat{B}} = \frac{\mathbf{B}}{\|B\|}
\label{eq:Bdir}
\end{equation}
and calculate $\psi$ with  
\begin{equation}
\psi = \frac{\pi}{2} + \arctan( \mathbf{\hat{B}} \cdot \hat{\phi},
\mathbf{\hat{B}} \cdot \hat{\theta})
\label{eq:psiangle}
\end{equation}
where the $\hat{\phi}$ and $\hat{\theta}$ are the directional vectors of the
HEALPix coordinate directions. The angle  between the magnetic field and
the plane-of-the-sky (POS) is
\begin{equation}
\cos^2 \gamma = 1 - (\mathbf{\hat{B}} \cdot \hat{D})^2,
\label{eq:cos2gamma}
\end{equation}
where $\hat{D}$ is the direction of the LOS. Based on the
non-polarized emitted intensity of the cell,
$I_\mathrm{orig}$, we get the Stokes components
$I$, $Q$ and $U$ with
\begin{equation}
I = I_{\mathrm{orig}} \Bigg[ 1 - p_0 \Bigg(\cos^2 \gamma - \frac{2}{3}\Bigg) \Bigg]
\label{eq:Ipol}
\end{equation} 
\begin{equation}
Q = I_{\mathrm{orig}} p_0 \cos(2 \psi) \cos^2 \gamma 
\label{eq:Qpol}
\end{equation} 
\begin{equation}
U = I_{\mathrm{orig}} p_0 \sin(2 \psi) \cos^2 \gamma.
\label{eq:Upol}
\end{equation}
To match the polarization degrees observed in the \PIX\ 
 and \PXX\ we set $p_0 = 0.2$. 

From the integrated $Q$ and $U$ we can calculate the polarization fractions
$p$ and the polarization angle dispersion functions $\Sd$ over the whole sky. 
Simulated $p$ and $\Sd$ yield familiar measures of system properties and
allow comparison to previous studies. 
The polarization fraction is defined as
\begin{equation}
p = \frac{\sqrt{Q^2 + U^2}}{I}
\end{equation}
and the polarization angle dispersion function
\citep{FaGo2008,hildebrand2009,Planck2015XIX} as 
\begin{equation}
\Sd(\mathbf{r}, \mathbf{\delta}) = \sqrt{ \frac{1}{N} \sum_{i=1}^N \big[ 
\psi(\mathbf{r}) - \psi(\mathbf{r} + \mathbf{\delta}_i) \big]^2 },
\end{equation}
where $\psi(\mathbf{r})$ is the polarization angle in the given position in the 
sky $\mathbf{r}$ and $\psi(\mathbf{r} + \mathbf{\delta}_i)$ the polarization 
angle at a position displaced from the centre by the vector 
$\mathbf{\delta}_i$. The sum extends over pixels whose distances from the 
central pixel in the location $\mathbf{r}$ are between $\delta/2$ and 
$3\delta/2$. As in \PIX,
we calculate the dispersion function with: 
\begin{equation}
\psi(\mathbf{r}) - \psi(\mathbf{r} + \mathbf{\delta}_i) =  
\frac{1}{2} \arctan(Q_i U_r - Q_r U_i, Q_i Q_r + U_i U_r). 
\end{equation}
In addition, to be comparable with \blue{the analysis presented in} \PIX\ 
we apply $1\degree$
Gaussian smoothing to our simulated observations, and set $\delta = 30'$ for all
calculations of $\Sd$. 
This sets the widest distance between points in the annulus to be $1.5$ times
the size of the FWHM of the Gaussian filter. 
The size of the annulus is small because, according to \PIX\, the angle
dispersion function $\Sd$ gradually loses its coherence with increasing
$\delta$. \PIX\ also note that their measurements of $\Sd$ are not an artefact
of either the choice of $\delta$ itself or instrumental bias.
However, the small annulus will include some influence due to spatial correlations
induced by the $1\degree$ beam. The choice of $\delta$ is independent of our radiative
transfer model, and its primary justification is to compare our results with \PIX.

We have calculated our results with several integration distances $\Rmax$,
namely $\Rmax = 0.25$, $0.5$, $1$, $2$, $4$ kpc, to explore the effect of
depth.
A part of the emission observed in Planck XIX is received from distances 
larger than what our model is capable of exploring 
However, with this in mind, our motivation is to inform, with an independently 
and physically generated ISM
and magnetic field, how the impact of near and distant features might
contribute to the observed images.
We assume periodic boundary conditions in the horizontal
direction of the computational domain.
However, if the boundary is crossed in the vertical
direction, the integration stops at that point, as happens in the cases
$\Rmax = 2$ and $4$ kpc.
Note, for our analysis, we mask latitudes $|b|>45\degree$ from
our maps.
This more closely reflects the region of the sky where the Planck observations
are included in the analysis of \PIX. 
This is an arbitrary choice, as we are not limited by the signal-to-noise 
issues of \PIX. However, this has also other benefits.
It avoids the problem of calculating $\Sd$ near the poles and some of the issues of asymmetry, caused by the limited vertical extent of the domain, that would affect the comparison of  $\Rmax \leq 1$\,kpc and $\Rmax \geq 2$\,kpc.
To save space, we omit from the figures related to the case with
$\Rmax = 0.5$ kpc, but the results are similar to $\Rmax = 1$ kpc. 

%-----------------------------------------------------------------------------
\subsection{Combining data into 2D-histograms}\label{sec:haver}
%-----------------------------------------------------------------------------

%-----------------------------------------------------------------------------
   \begin{figure*}
   \centering
      \begin{subfigure}[t]{0.35\textwidth}
         \includegraphics[width = 0.9\hsize]{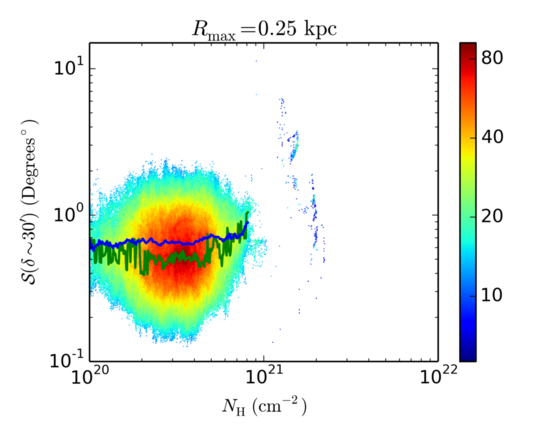}
      \end{subfigure}
      \begin{subfigure}[t]{0.4\textwidth}
         \includegraphics[trim=0.0cm 0.0cm 0.0cm 0.0cm,clip=true,width = \hsize]{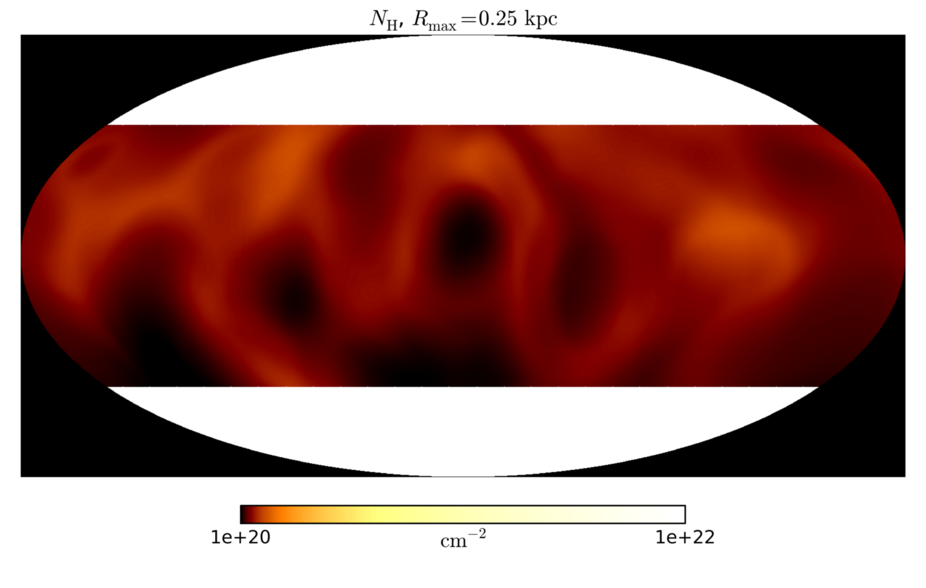}
      \end{subfigure}
      \begin{subfigure}[b]{0.35\textwidth}
         \includegraphics[width = 0.9\hsize]{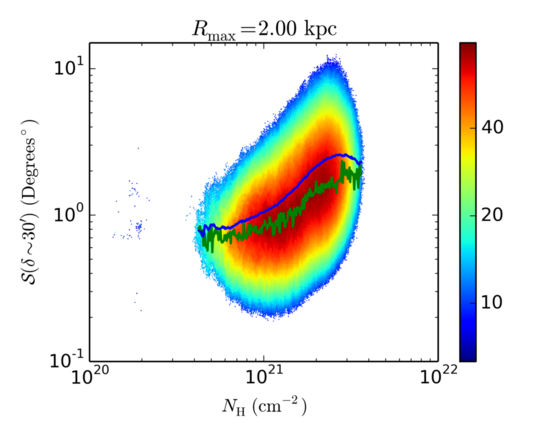}
      \end{subfigure}
      \begin{subfigure}[b]{0.4\textwidth}
         \includegraphics[trim=0.0cm 0.0cm 0.0cm 0.0cm,clip=true,width = \hsize]{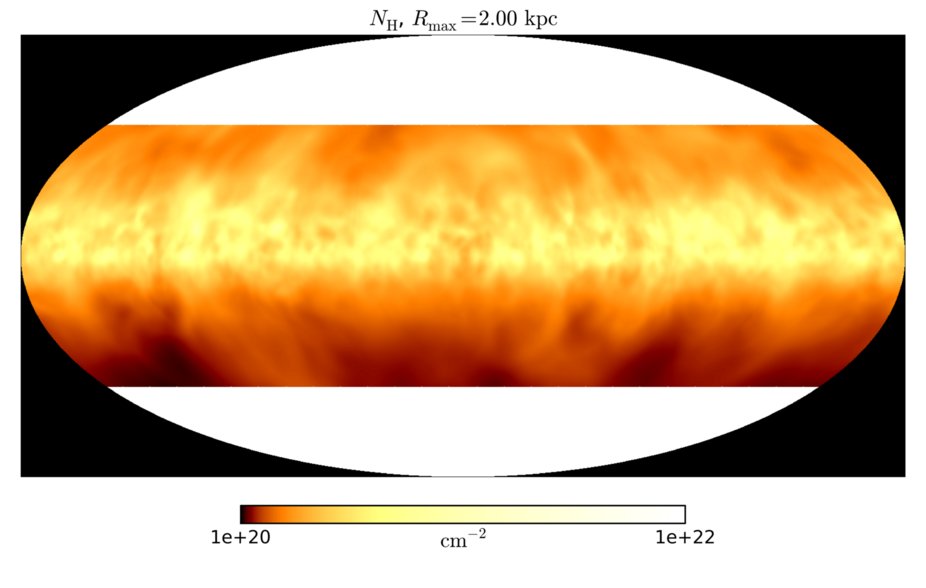}
      \end{subfigure}
      \begin{subfigure}[b]{0.35\textwidth}
         \includegraphics[width = 0.9\hsize]{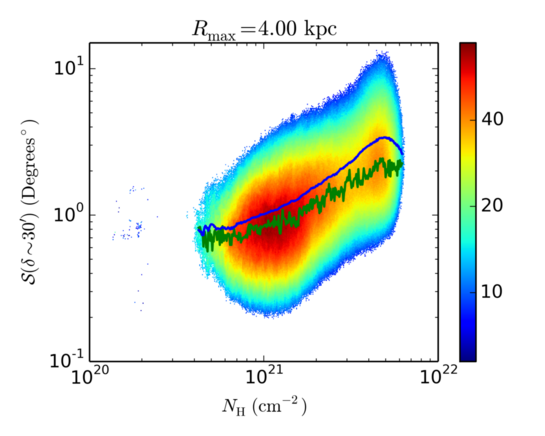}
      \end{subfigure}
      \begin{subfigure}[b]{0.4\textwidth}
         \includegraphics[trim=0.0cm 0.0cm 0.0cm 0.0cm,clip=true,width = \hsize]{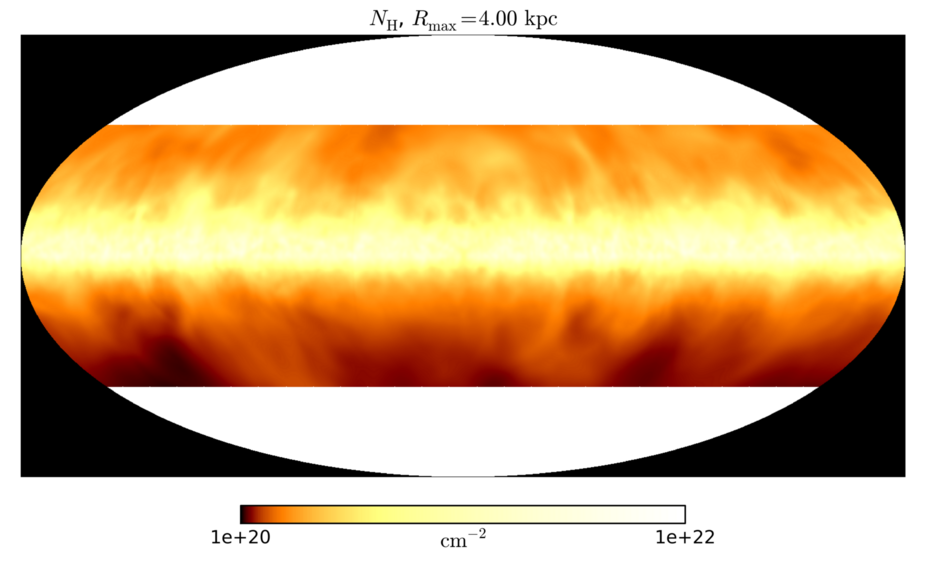}
      \end{subfigure}
      \caption{{\it Left:} Joint 2D-histograms of polarization angle
               dispersion $\Sd$ and column density $\NH$.
               The green lines follow the maximum $\cbin$ as a function of $\NH$, 
               and the blue lines present the weighted mean of $\Sd$. 
               {\it Right:} Sample maps of column density.
	       For the case $\Rmax = 1 \, \mathrm{kpc}$, see Fig. \ref{fig:hist_NS_flux}.  
               }
         \label{fig:pol_Nmap}
         \label{fig:hist_NS}
   \end{figure*}
%-----------------------------------------------------------------------------

Many of the results in this paper are represented using 
joint 2D-histograms. This allows us both to compare our results directly to the
analysis in \PIX\ 
and \PXX, and separate the
snapshot and point-of-view (POV)
specific details from the general large-scale behaviour. 

For 2D-histograms, we combine radiative transfer simulations from all
12 snapshots and for 5 different POV positions in the
approximate
midplane, specifically from the points
$(0.5,  0.5,  1.0)$ kpc, 
$(0.25, 0.5,  1.0)$ kpc, 
$(0.5,  0.25, 1.0)$ kpc, 
$(0.75, 0.5,  1.0)$ kpc and  
$(0.5,  0.75, 1.0)$ kpc,
under the assumption that the use of periodic boundary conditions in $x$- and $y$-directions
should give a reasonable representation of the ISM in the neighbourhood of the Solar System. 

The data is combined as follows. 
From each SOC-generated map, $i$=1-60, separate 2D-histograms are constructed over a common
set of $512 \times 512$ bins for each comparison, specifically
$\cbini(\NH,p)$, $\cbini(\NH,\Sd)$, $\cbini(p,\Sd)$. Here $\cbini$
represents the number of counts in a single bin calculated from a single
synthetic observation. 
Then, assuming $\cbini$ to be lognormally distributed
between separate 2D-histograms, we calculate the mean 
and standard deviation over $i$, 
\begin{equation}
\log \cbin = \langle \log \cbini \rangle, \quad
\sigma_{\log\cbin} = \sigma(\log \cbini). 
\end{equation}

We include only the bins $\cbin$ for which $\cbini > 0$ for all $i$. 
Second, to reduce additional noise, we mask out the elements where
\begin{equation}
 \exp(\log \cbin - \sigma_{\log\cbin}) < 5, 
\end{equation}
yielding the composite 2D-histograms, as presented in Fig.\,\ref{fig:pol_Nmap}
and subsequent 2D-histogram figures.
Combined 2D-histograms, summing counts over many snapshots and POV,
contain the patterns most consistent across the individual 2D-histograms.
Despite masking low count noise, many low-frequency outliers persist.

The various simulated observations may each exhibit distinct features worthy
of further analysis in the future. However, in this study we
focus on the general behaviour of the system. 
For example, the 2D-histograms only include pixel-by-pixel numerical 
values. Therefore, such analysis cannot regard the filament-like 
\textit{shapes} visible in $\Sd$-maps (See Figs. \ref{fig:local}, 
\ref{fig:pol_disp} and \ref{fig:pol_disp_flux}).

%-----------------------------------------------------------------------------
\section{Properties of synthetic polarization observations}\label{sec:results}
%-----------------------------------------------------------------------------

Here we present our results and analysis based on the synthetic observations 
obtained using SOC with polarization tools on top of our MHD simulations.
For our analysis, we use the maps of column densities $\NH$, polarization
fractions $p$ and polarization angle dispersions $\Sd$. 
As discussed in Sect.\,\ref{sect:mhdsim}, we take (in the figures) as the reference case the 1.7\,Gyr 
snapshot, where the observer is situated in the centre of the computational domain.  
We opt for projection onto a sphere as the most appropriate 
observational reference frame. 
With the domain spanning over 1\,kpc in each direction, it is too large to be
treated as a single observation along a Cartesian coordinate. 
Thus, projection onto a sphere is the most realistic observational frame, and
especially so when nearby regions are concerned. 
The all-sky point of view affords the best comparison with Planck on the
scales relevant to the MHD model. 
All of the maps of synthetic observations over the whole sky are presented
using Mollweide projection.

\blue{\subsection{Simulated column density}\label{sec:NH}}

The maps of column density resulting from integrating over each distance, up to
$\Rmax=0.25,\,2$ and 4\,kpc from the observer, are shown in Fig.\,\ref{fig:pol_Nmap},
right panels.
The dependence of $\NH$ on $\Rmax$ is immediately apparent.
The densest regions are situated near the midplane, as would be expected
from horizontal averaging of the gas density in the MHD model represented in
Fig.\,\ref{fig:rhoT}.
However, this is not visible on the map from $\Rmax=0.25$\,kpc, but clear for
the higher $\Rmax$.
The vertical anisotropy in the latter reflects
the temporal upward shift of the disk centre of mass, evident from
Fig.\,\ref{fig:rhoT}. 
Also, density should be near isotropic in the latitudinal
direction, because we do not model the galaxy's central bulge nor spiral arms.
In a single snapshot, however, local SN bubbles or superbubbles (merging
multiple SN remnants) may inject significant anisotropy into the overall
density profile.
This is clearly seen only for the $\Rmax < 1$\,kpc, with bubbles
apparent on several locations. 
For the higher lengths of integration the influence of features near to the 
observer is almost negligible.
As the combination of nearby and far-away emission will be present in both
real and synthetic observations, exploring the effect of distance to the
observations is called for.

Fig.\,\ref{fig:pol_Nmap}, 
left panels, 
show joint 2D-histograms of polarization angle dispersion $\Sd$ and
column density $\NH$ for $\Rmax = 0.25,\,2$ and 4\,kpc.
At 0.25\,kpc, $\NH$ is clustered around only
$1.5\cdot10^{20}$\,cm$^{-2}$ and $\Sd<1\degree$, while its corresponding $\NH$
map, dominated by local gas structure, is without dense clouds.
This is inconsistent with \PIX\ and indicates that the integration length is too
short.
For higher $\Rmax$, $\NH$ is typically above $10^{21}$\,cm$^{-2}$ and 
consistent
with the observed column density featured in \PIX, Fig.\,24 upper panel.
$\Sd$ also increases above 1, but not to the level obtained by \PIX.
The high $\Rmax$ maps of $\NH$ also capture the presence of these clouds,
particularly evident near the midplane.

On these joint 2D-histograms are also plotted lines of maximum $\cbin$ (green)
and weighted mean (blue) polarization fractions as functions of $\NH$.
For $\Rmax\gtrsim1$\,kpc the trend has a positive correlation with $\NH$,
in disagreement with \PIX\ Fig.\,24.
There is no record in \PXX\ on how this relationship applies to their MHD model.
Comparing our synthetic observations with \PIX\ Fig.\,24, there is a peak in
the weighted mean profile for high column density dust.
This corresponds to the molecular clouds, which are not resolved in the MHD
model.
In Fig.\,\ref{fig:pol_Nmap} ($\Rmax>1$\,kpc) the high column density range,
representing the warm and cold ISM, is a good match for the \PIX\ data.
In the \PIX\ data, however, the weighted mean remains as high or even increases
at lower densities, while our simulations show a correlation ${\cal S}\propto\NH^{0.5}$
in this range of column densities, being in obvious disagreement.

Let us try to understand this discrepancy by assuming that the small-scale
  scatter seen through $\cal{S}$ is a result of the underlying turbulent diffusion,
  taking the action of smoothing the flow. If this was the case, one would expect
  the higher viscosity regions smooth out structures, also those seen in $\cal{S}$, more efficiently,
  that is a relation ${\cal{S}}\propto\nu_t^{-1}$ would be expected, where $\nu_t$
  is a turbulent diffusion coefficient related to turbulent mixing,
not necessarily to any dependence assumed in viscosity scheme used in the model.
As explained in Sect.\,\ref{sect:mhdsim},
gas viscosity is set as $\nu\propto T^{0.5}$ so as to resolve flows
in the hot gas of the MHD model.
The ISM is modelled as an ideal gas, with all phases in approximate
statistical pressure balance, so this is statistically equivalent to 
$\nu\propto\rho^{-0.5}$.
For the radiative transfer calculations, the dust density is assumed proportional to the
ISM gas density.
If higher viscosity in the hot gas tends more proportionally to smooth
small-scale structures in the flow, we might expect the dispersion
${\cal S}\propto\NH^{0.5}$,
which we indeed see for the low-density hot gas.
This implies that the diffusion in the hot gas reflects the dependence input
through the viscosity scheme, while the one for the cold and warm gas does
not do so, but is somewhat steeper than expected from this simple argumentation.

We have no reliable method to determine observationally what is the true
turbulent viscosity in the ISM.
Some elaborate and approximate methods make this possible within numerical
experiments \citep[see, e.g.,][]{KGVS17}.
Our simple hypothesis presented above could be tested by applying the same analysis in this
paper to MHD simulations with alternative prescriptions for viscosity, to
exclude a relationship between the weighted mean profile and the viscosity, as
appears to exist here.
However, if MHD models consistently display such a dependence on viscosity
$\Sd\propto\nu^{-1}$
for the weighted mean $\cal{S}$, then we might be able to infer something 
about the actual turbulent viscosity in the ISM from the \PIX\ profiles.
Using \PIX\ with the weak negative correlation in between the weighted mean values
  of $\cal{S}$ and $N_{\rm H}$ would
imply $\nu\propto T^{-\lambda}$ for the ionized ISM, with $0<\lambda\ll1$.
This is quite unlike, e.g., Spitzer molecular viscosity
$\nu\propto \rho^{-1}T^{2.5}$. 
The scale of such simulations place that investigation beyond the scope of
the present study, and of course there are many complicating factors which would permit alternative explanations of this trend.

At the lowest $\Rmax$ the visible cloud structure \blue{in the $N_H$ map} 
is \blue{directly} tracing our MHD model.
With increasing $\Rmax$ smaller details appear in the maps, which are
\blue{the effects of projection from more remote sources}.
Hence, the observed results are inevitably a combination of both the MHD model
volume and its projection through a mixture of features along the LOS. 
By assuming periodicity in the horizontal direction, we can reach higher
$\Rmax$ than the physical scale of our MHD model. However, with too high $\Rmax$
the periodicity creates a mirror gallery effect of repeating patterns near the
$x$- and $y$-axes of the MHD model. 
To avoid this, it is reasonable to restrict the highest integration
distance to $\Rmax = 2$ kpc, or for the general case, $\Rmax\lesssim2L_x$,
where $L_x$ is the horizontal extent of the MHD model.

%-----------------------------------------------------------------------------
\subsection{Polarization fraction across the galactic plane}\label{sec:pol}
%-----------------------------------------------------------------------------

%-----------------------------------------------------------------------------
   \begin{figure*}
   \centering
      \begin{subfigure}[t]{0.35\textwidth}
         \includegraphics[width = 0.9\hsize]{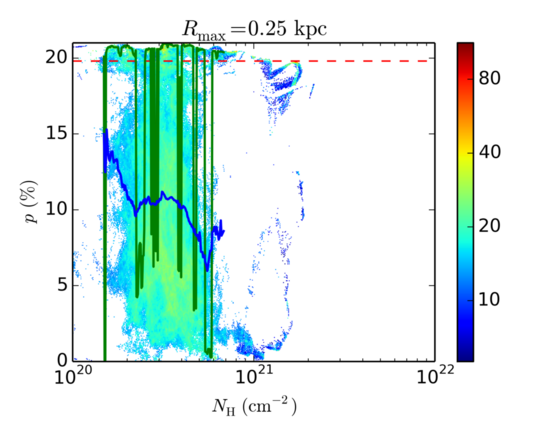}
      \end{subfigure}
      \begin{subfigure}[t]{0.4\textwidth}
         \includegraphics[width = \hsize]{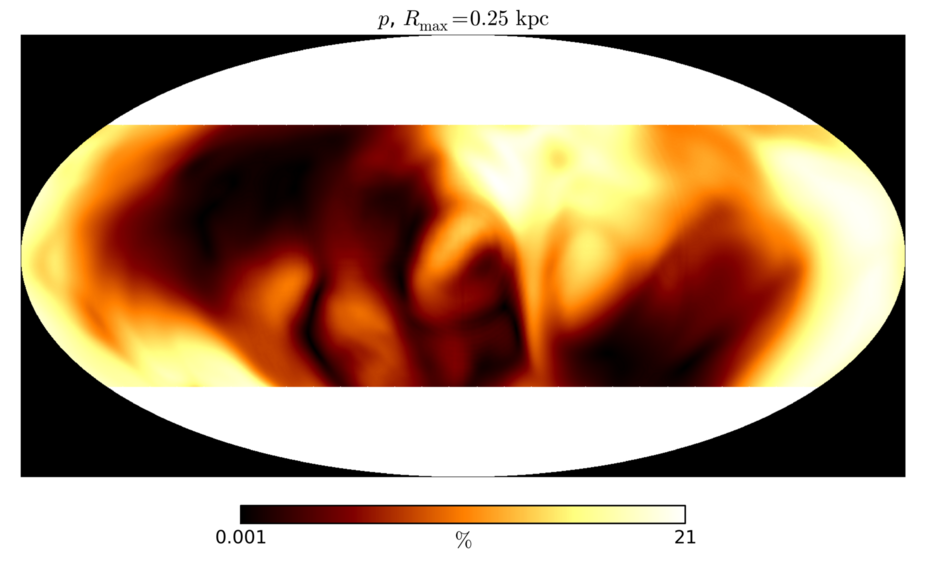}
      \end{subfigure}
      \begin{subfigure}[b]{0.35\textwidth}
         \includegraphics[width = 0.9\hsize]{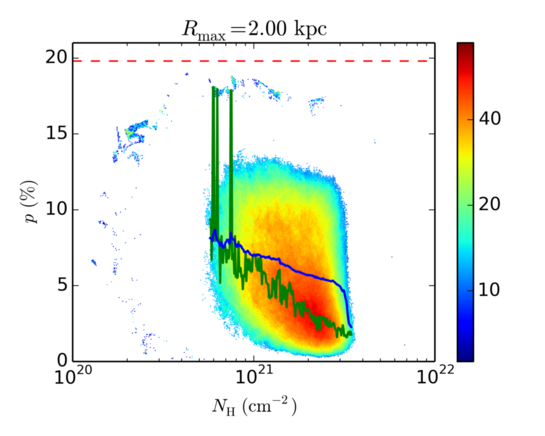}
      \end{subfigure}
      \begin{subfigure}[b]{0.4\textwidth}
         \includegraphics[width = \hsize]{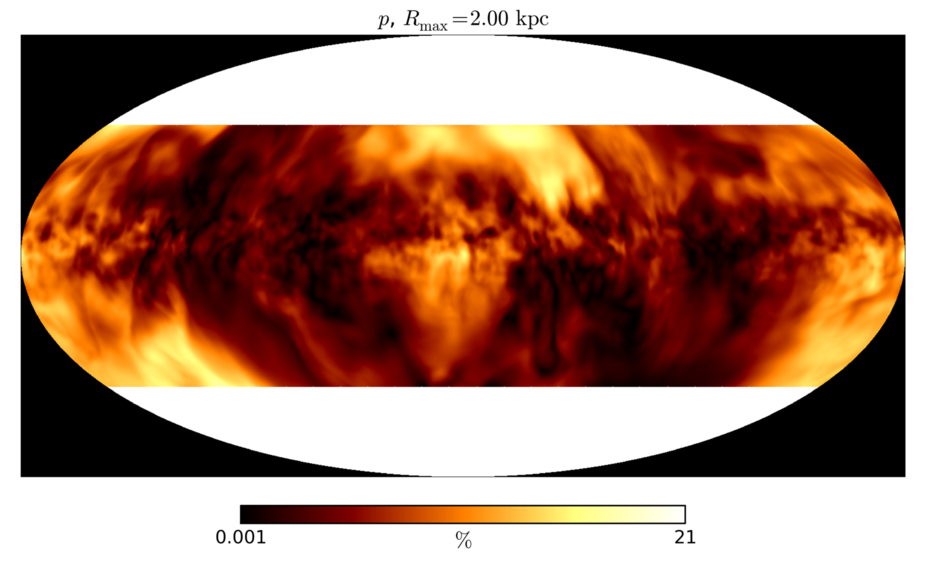}
      \end{subfigure}
      \begin{subfigure}[b]{0.35\textwidth}
         \includegraphics[width = 0.9\hsize]{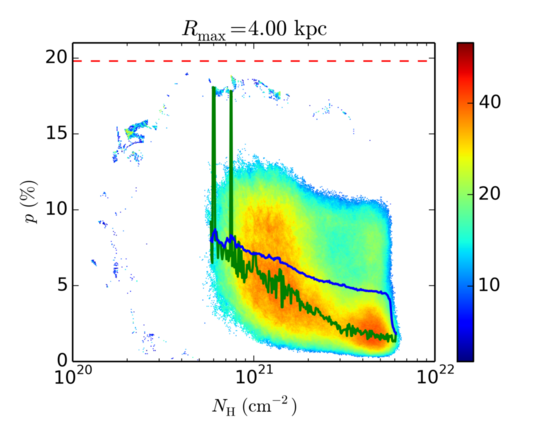}
      \end{subfigure}
      \begin{subfigure}[b]{0.4\textwidth}
         \includegraphics[width = \hsize]{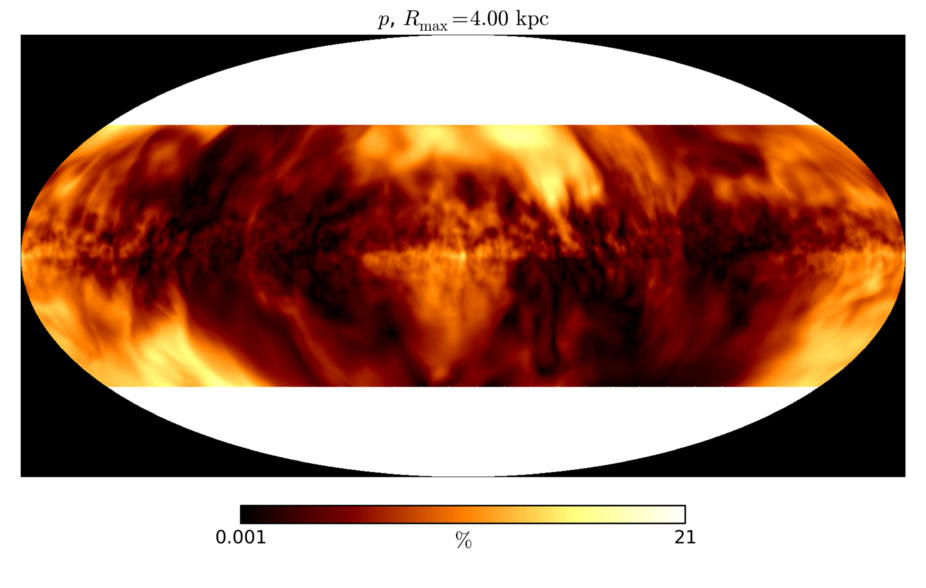}
      \end{subfigure}
      \caption{{\it Left:} Joint 2D-histograms of polarization fraction $p$ and
         column density $\NH$.
         Red dashed lines show the location of $p_\mathrm{max} = 19.8\%$
         from \PIX.
         The green lines follow the maximum $\cbin$ as a
         function of $\NH$, and the blue lines present the weighted mean of $p$. 
         {\it Right:} Sample maps of $p$.
         The polarization fractions are weakened in the galactic midplane,
         which is seen in the 2D-histograms as a growing distribution of low
         $p$ at high $\NH$.
         For $\Rmax = 1 \, \mathrm{kpc}$, see
         Figs.\,\ref{fig:pol_pmap_flux} and \ref{fig:hist_Np_flux}.
               }
         \label{fig:pol_pmap}
         \label{fig:hist_Np}
   \end{figure*}
%-----------------------------------------------------------------------------

%-----------------------------------------------------------------------------
   \begin{figure*}
   \centering
   \end{figure*}
   \begin{figure*}
   \centering
      \begin{subfigure}[t]{0.35\textwidth}
         \includegraphics[width = 0.9\hsize]{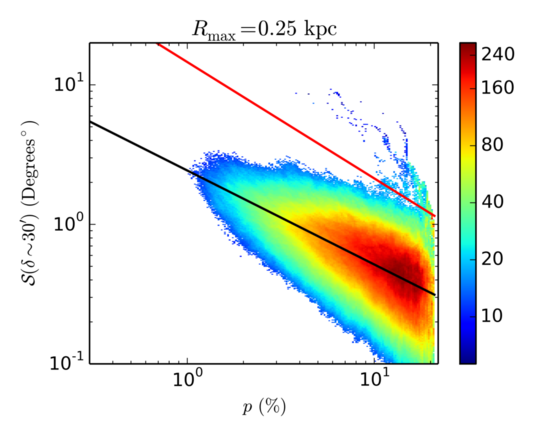}
      \end{subfigure}
      \begin{subfigure}[t]{0.4\textwidth}
         \includegraphics[width = \hsize]{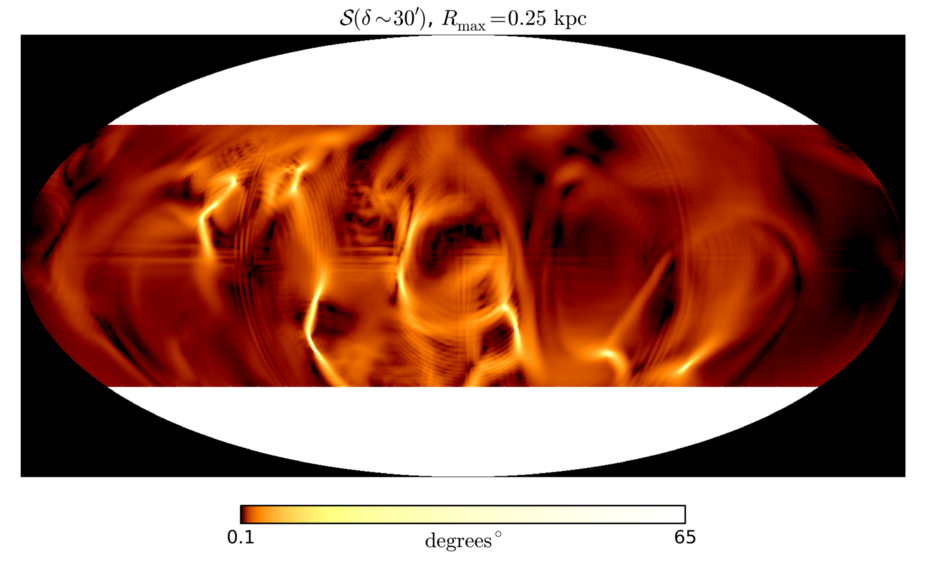}
      \end{subfigure}
      \begin{subfigure}[b]{0.35\textwidth}
         \includegraphics[width = 0.9\hsize]{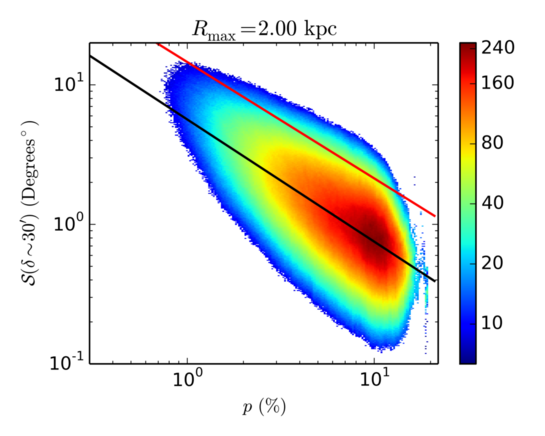}
      \end{subfigure}
      \begin{subfigure}[t]{0.4\textwidth}
         \includegraphics[width = \hsize]{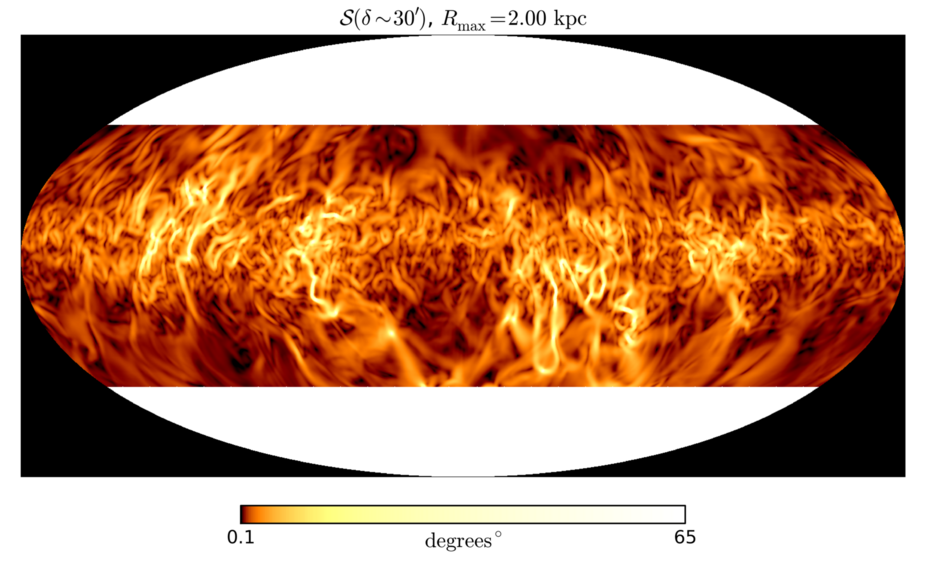}
      \end{subfigure}
      \begin{subfigure}[b]{0.35\textwidth}
         \includegraphics[width = 0.9\hsize]{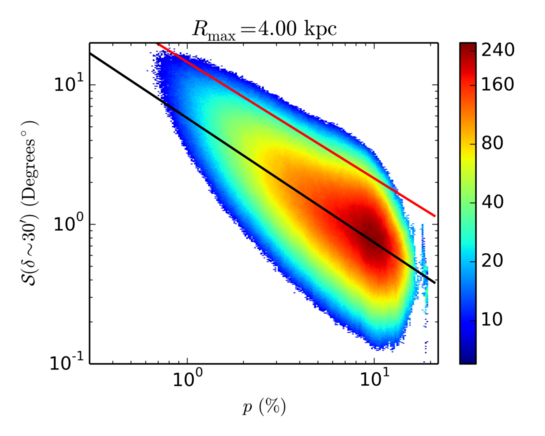}
      \end{subfigure}
      \begin{subfigure}[b]{0.4\textwidth}
         \includegraphics[width = \hsize]{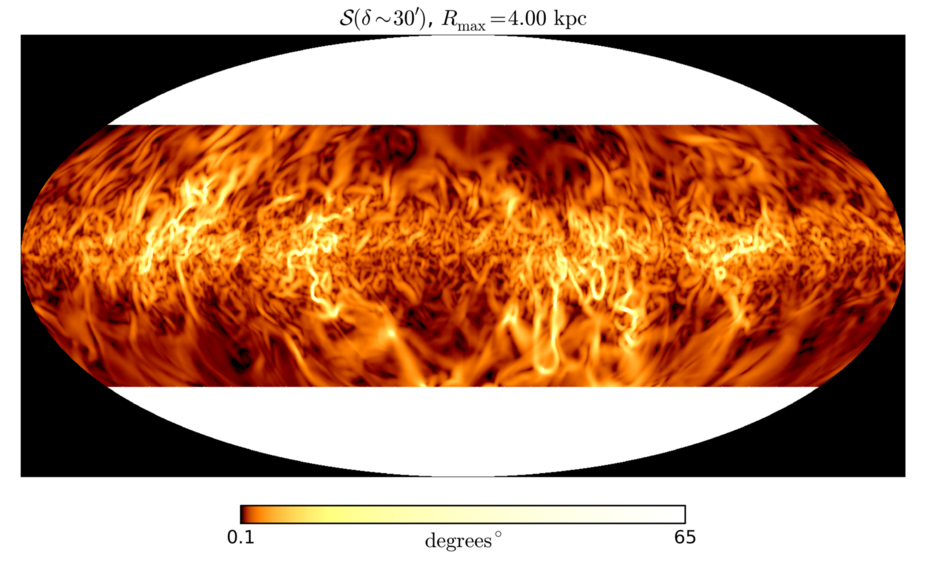}
      \end{subfigure}
      \caption{{\it Left:} Joint 2D-histograms of angle dispersion and
         polarization fraction, with the effect of increasing $\Rmax$.
         The red line depicts the fit to the observations as presented in \PIX,
         $\log_{10} \Sd = \alpha \log_{10}p + \beta$. 
	 The black lines are the best fits to our simulated data, to
         which $\alpha$ and $\beta$ are given in the Table\,\ref{tbl:powerfit}. 
         {\it Right:} Sample maps of polarization angle dispersion $\Sd$ 
         with corresponding $\Rmax$.
             }
         \label{fig:pol_disp}
         \label{fig:hist_pS}
   \end{figure*}
%-----------------------------------------------------------------------------

For each $\Rmax$ we construct joint 2D-histograms of polarization
fraction $p$ and $\NH$, presented in Fig.\,\ref{fig:hist_Np}, 
left panels, 
and the maps of $p$,
right panels.
With $\Rmax=0.25$\,kpc there is a broad range for $p$ values. 
The distribution is independent of the
$\NH$ values, which in this case are much lower than in the \PIX\ observations.
The mapping of $p$ can be seen to correlate over large smooth regions. 
There is negligible depolarization, with the highest $\cbin$ appearing
on the scale of $p_0 = 20\%$.

With $\Rmax=2$\,kpc depolarization is stronger, particularly near the midplane,
the map of $p$ being an excellent analogue for \PIX, Fig.\,6.
Its 2D-histogram is also a better match with \PIX, Fig.\,19, although
$\NH>10^{22}$\,cm$^{-2}$ are absent and there are few values with $p>15\%$.
The high column density may relate to features not modelled with the MHD,
such as the central bulge, spiral arms, and molecular cloud properties
requiring higher resolution to be resolved. 
Some of the absence of higher $p$ is due to the masking of bins which lack signal across 
all synthetic maps cobined to the 2D-histogram in question.
In addition, the fraction of the \PIX\ data, which has $p>15\%$, is concentrated into the 
molecular cloud complexes, where high polarization fractions can be found, which are among 
the features not resolved in our MHD models.

The increased frequency of $p>5\%$ with $\Rmax=4$\,kpc is more like the
\PIX\ results, but the high number of points where 
$p<5\%$ and $\NH\simeq5\cdot10^{21}$\,cm$^{-2}$ is not consistent with \PIX, and an indication 
that oversampling the same periodic domain near the midplane is 
distorting the distribution.
To improve the range of column densities in a physically consistent manner, 
one should increase the horizontal extent of the MHD models.
Also, increasing the MHD model resolution would improve both the proportion of
high column densities and high polarization fractions, 
but this would be computationally demanding in the context of SN-driven
MHD turbulence that is capable of producing self-consistent small- and
large-scale dynamos.

Regardless of these current limitations, it is interesting to consider how the 
distances influence polarization and 
depolarization.
In the framework of the radiative transfer calculations, the mechanism is
easy to understand.
Along the LOS, individual cells in the grid produce positive and
negative contributions in $Q$ and $U$, depending on the local magnetic field direction.
For an incoherent magnetic field, the sign of $Q$ and $U$ fluctuates, with
similar magnitude, such that over very long distances in an
optically thin medium their averages approach zero.
This is comparable to the analysis of \citet{houde2009}, where they note
that multiple independent turbulent cells along the line of sight will 
make the apparent fluctuation to approach zero. 
However, the mean field 
also varies in direction when exploring kiloparsec scales which will also contribute to the perceived effect. 
In the presence of a directionally coherent magnetic field, its
direction will dominate the $Q$ and $U$ over long distances.
However, the magnetic fields are highly turbulent throughout the MHD domain, so we
expect depolarization to increase with $\Rmax$.

Near the midplane the depolarization effect is strongly proportional to
integration length with $\Rmax\geq2$\,kpc (Fig.\,\ref{fig:hist_Np}, lower right
panels), corresponding high $\NH$.
This is also consistent with the above explanation.
Near the midplane with large $\Rmax$ it is possible to integrate
polarization over a longer distance, unlimited by the upper or lower boundaries
of the computational domain.
Therefore, there is more influence on $p$ from the depolarizing effect of
the fluctuations, most evident near the midplane. 
The role of magnetic field fluctuations inducing a broad spread of $p$ has
also been suggested by \PIX\ and 
\PXX\,  
and our results support this idea.

Due to the averaging and masking method presented in
Sect.\,\ref{sec:haver} some of the outlying values persist as halos in the
2D-histograms of Fig.\,\ref{fig:hist_Np} (also Fig.\,\ref{fig:hist_Np_flux}), 
but with higher sampling rates, the masked points between the bulk data
and halo could be restored.
Also, the gradual loss of alignment of the dust grains within radiation shielded
dense clouds can decrease polarization, which is also considered by \PIX\ to be 
an effect influencing their observed depolarization with $\NH  \geq 10^{22}$\,cm$^{-2}$.
We set the strength of alignment proportional to $p_0$ in
Eqs.\,\ref{eq:Ipol}-\ref{eq:Upol}, neglecting any effects of such shielding.

As our methods resemble those of \PXX, some dicussion is called for, 
although we cannot make a direct comparison of their MHD model
with our results. 
They consider scales well below our 4\,pc grid resolution.
What is relevant, is that the structure of the flow and the magnetic field in 
our MHD model is naturally driven by the forces on the scales of SN remnants
cascading to the lowest turbulent eddies we can resolve. 
It is reasonable to expect that the turbulent structure would extend to 
lower scales, until new physical processes become active.

\PXX\ compare their MHD results with observations of the
Chamaeleon-Musca and Ophiuchus fields.
For their MHD simulation, they view the domain with respect to the mean field 
as POS, LOS and 50/50, confirming that a high polarization fraction is indicative of a
strong POS component to the field.
This is evidence that the magnetic field has a strong POS component in 
Chamaeleon-Musca, while in Ophiuchus the field is more aligned along the LOS, an
interpretation consistent with \citet{PlanckXXXV16}, their Fig.\,3,
which shows a more consistently ordered POS field for Chamaeleon-Musca compared
to that of Ophiuchus.

For all three cases examined in \PXX, the values in the polarisation fraction distribution are low compared
to either Chamaeleon-Musca or Ophiuchus.
It may be that the random component, and hence depolarization, in the MHD 
models was too strong.
Alternatively, the forcing mechanism they use induced a Gaussian distribution to the magnetic field, while
our analysis, with SN-driven turbulence, indicates it to have more exponential 
distribution, which could influence the efficiency of the depolarization 
(see Sect.\,\ref{sec:poldisp}).
With respect to the mean field component
\citet[][Ch.\,9 Fig.\,9.12]{Gent2012}, find that the magnetic fields in the
cold filamentary regions formed by SN driven turbulence are more regular
than the ISM as a whole and that they are strongly aligned with the ambient
warm ISM, in which they are embedded.
This is in contrast to the observations of \citet{Planck2016XXXII,Planck2016XXXIII}, who
found the orientation of the filamentary structure of the most dense 
molecular clouds to be perpendicular to the mean magnetic field.
Here, gravitational collapse and runaway thermal instabilities leading to the
formation of such high density structures may be critical,
which are absent on the scales considered in our MHD model.

%-----------------------------------------------------------------------------
\subsection{Polarization angle dispersion}\label{sec:poldisp}
%-----------------------------------------------------------------------------

Looking at the maps of polarization angle dispersion $\Sd$ presented in 
Fig.\,\ref{fig:pol_disp} (right panels),
we observe filamentary structures ($\Sd$-filaments) similar in appearance to 
\PIX\ Fig.\,12.
However, for $\Rmax=0.25$\,kpc these are very large scale structures, which span
the full range of examined galactic latitudes in some locations and are much thicker than the \PIX\
observations.
With increasing $\Rmax$ the $\Sd$-filament structure resembles  quite well
\PIX\ Fig.\,12, becoming tangled and fragmented, increasing in number and having
more wiggles.
\PXX\ report filament-like maps of $\Sd$ from their synthetic observations.
However, due to the differences in scale, resolution and forcing mechanism 
behind the MHD model used in \PXX\ and this study, an effective qualitative comparison is not reasonable. 

\begin{table}[h]
\centering
\begin{tabular}{|r|r|r|r|l|}
\hline 
$\Rmax$ (kpc)& $\alpha$ & $\beta$& $\langle \Sd/\Sd_\mathrm{Planck} \rangle$ & Notes \\ 
\hline
--   & $\mathbf{-0.834}$ & $\mathbf{-0.504}$& $100\%$ & Planck \\
0.25 & $-0.673$ & $-0.961$ & $24\%$ &  \\
0.5  & $-0.742$ & $-0.898$ & $32\%$ &  \\
1.0  & $-0.789$ & $-0.920$ & $34\%$ &  \\
1.0  & $-0.934$ & $-1.020$ & $39\%$ & $2 \cdot \mathbf{b}$ \\
2.0  & $-0.876$ & $-0.999$ & $35\%$ &  \\
4.0  & $-0.892$ & $-1.021$ & $35\%$ &  \\
\hline
\end{tabular}
\caption{Fits to $\log_{10} \Sd = \alpha \log_{10}p + \beta$ for
each joint 2D-histogram of $\Sd$ and $p$. The observed values are from the best fit of
\PIX.}
\label{tbl:powerfit}
\end{table}

In Fig.\,\ref{fig:hist_pS} 
(left panels)
the joint 2D-histograms of $\Sd$ and $p$ show some agreement with
Fig.\,23 of \PIX.
Angular dispersion is inversely proportional to polarization fraction, and may 
be approximated by
\begin{equation} 
\label{eq:logS}
         \log_{10} \Sd = \alpha \log_{10}p + \beta. 
\end{equation} 
The 2D-histograms show our best fits (black lines) for Eq.\,\eqref{eq:logS} and the
fit from \PIX\ (red lines).
The parameters are summarised in Table\,\ref{tbl:powerfit}.
For $\Rmax$ at 1\,kpc and above, the slope $\alpha$ is near to
the \PIX\ fit, but by having smaller intercept $\beta$ our 2D-histograms are shifted to lower $\Sd$.
There may be a case for inferring that the optimal integration length is about 
$\Rmax = 1.5$\,kpc, if we are looking to match $\alpha$ for the \PIX\ data.
Repeating the analysis for an MHD domain with increased horizontal extent and/or
increased resolution might indicate whether this is a robust physical 
relationship between the simulated and observed measurements.
The ratio 
$\Sd/\Sd_\mathrm{Planck} = p^{\alpha - \alpha_\mathrm{Planck}}
10^{\beta - \beta_\mathrm{Planck}}$ is averaged and listed in 
Table\,\ref{tbl:powerfit} for each $\Rmax$.

The dispersion values in our simulations increase towards \PIX\ as
$\Rmax$ increases, but plateau at a ratio of $35\%$ for $\Rmax\gtrsim1$\,kpc.
The increase to 39\%, when artificially strengthening 
the amplitudes of the small-scale fluctuations of the magnetic field,
supports the view that the low simulated value for $\beta$ is in part
due to insufficient small-scale field.
The low values of $\Sd$ in our MHD model may be attributed to truncated
resolution at the grid scale of 4\,pc.

\begin{figure}
   \centering
      \begin{subfigure}[b]{0.4\textwidth}
         \includegraphics[width = \hsize]{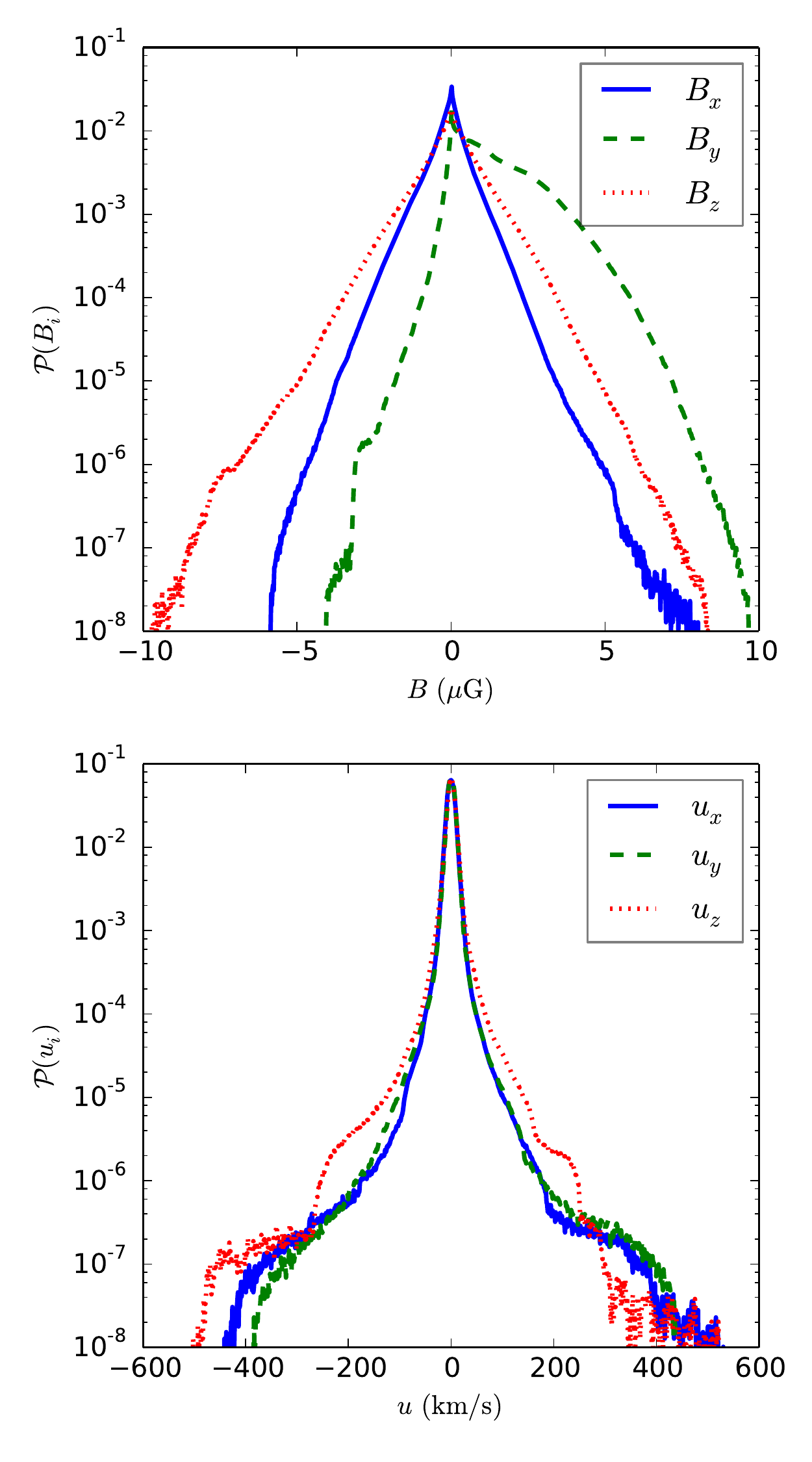}
      \end{subfigure}
      \caption{
      \blue{
           Probability density function combined from all 11 snapshots for 
           the components of $\mathbf{B}$ and $\mathbf{u}$.
               }}
         \label{fig:Bpdf}
   \end{figure}

\begin{figure*}
   \centering
      \begin{subfigure}[b]{1.0\textwidth}
         \includegraphics[width = \hsize]{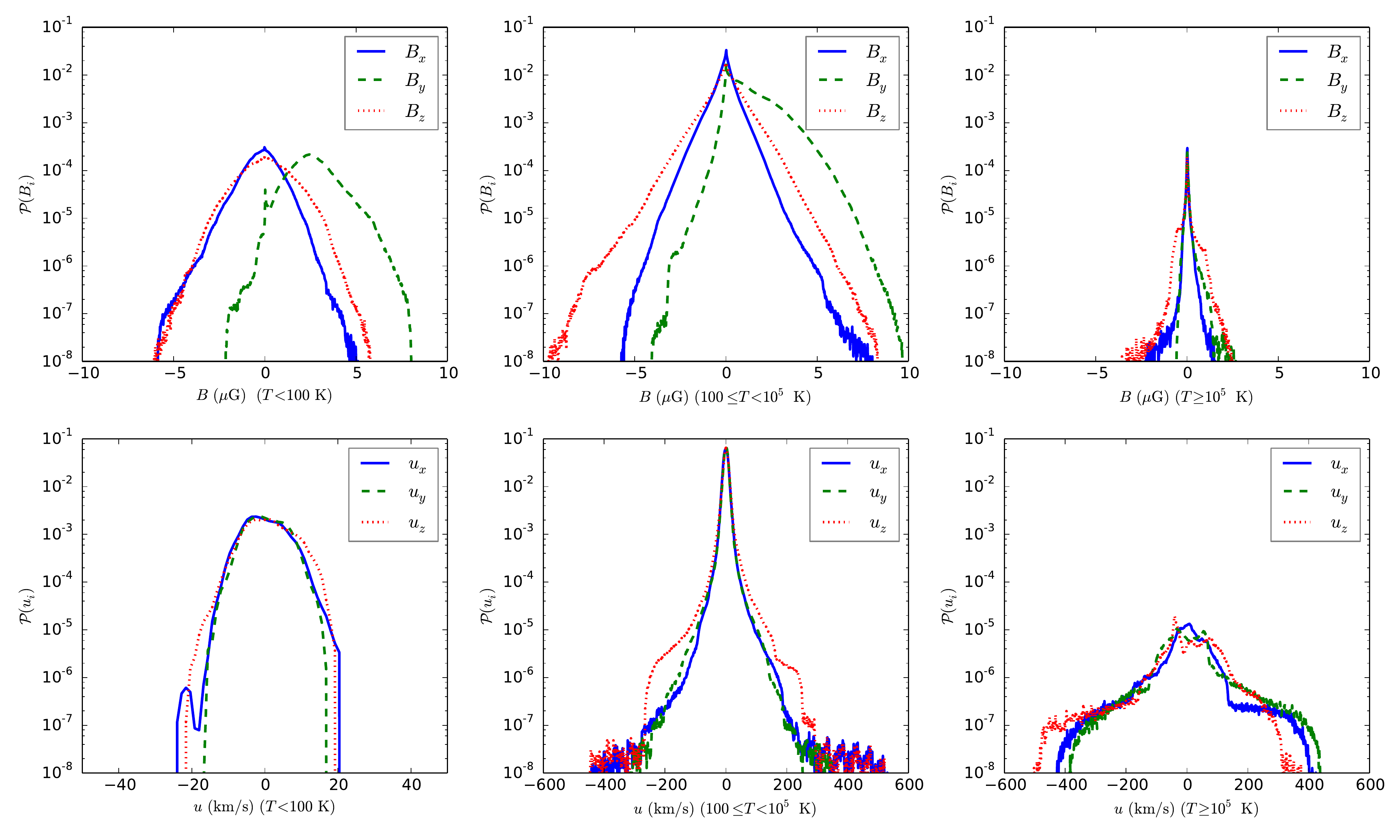}
      \end{subfigure}
      \caption{
           Probability density function combined from all 11 snapshots for 
           the components of $\mathbf{B}$ and $\mathbf{u}$ with cold, warm 
           and hot components separated.
               }
         \label{fig:Bpdf_hwc}
   \end{figure*}

To understand the distribution of velocity and magnetic fields in our MHD model data,
we present probability density distribution functions (PDFs) of the both variables in
Figs.\,\ref{fig:Bpdf} and \ref{fig:Bpdf_hwc}.
The PDFs are calculated over all MHD model cells from all snapshots where each cell, being 
of same volume, has an equal weight. 
In Fig.\,\ref{fig:Bpdf}, upper panel the PDF is exponential for
$B_x$ ($B_y$) and $B_z$, while $B_y$ ($B_x$) is skewed by the global shear.
Fig.\,\ref{fig:Bpdf}, lower panel, shows the PDFs for the velocity components, 
where within the velocity range $\pm400$\,km\,s$^{-1}$, the PDFs reflect ISM
physics as resolved by the model. 
We evolve SN remnants only from the latter stages of the Sedov-Taylor 
phase, however, so this is exhibited in the truncated PDF for
$|u|\gtrsim400$\,km\,s$^{-1}$.
%%%%%%%%%%%%%%%%%%%%%%%%%%%%%%%%%%%%%%%%%%%%%%%%%%%

The spiked PDF displayed in Fig.\,\ref{fig:Bpdf} for the magnetic and velocity 
field components arise from the physics of repetitive shock-driven turbulence,
independent of the model and resolution. 
Few authors have discussed this property, but a
 similar distribution for the velocity profile is illustrated in \citet[][Fig. 3.13]{Gressel2009}.
In their Fig.\,9 \citet[][]{MBKA05}, using approximately 1.5\,pc resolution,
 show similar PDF profiles for the divergence
of the velocity field, also supporting this physical interpretation of the
turbulent structure of the ISM.

The multiphase medium plays a role in this distribution. In
Fig.\,\ref{fig:Bpdf_hwc} we divide PDFs into three components, corresponding with cold, warm and
hot medium ($T<100$\,K, $100 \leq T < 10^5$\,K and $T>10^5$\,K respectively). 
The Fig.\,\ref{fig:Bpdf} velocity profile mainly exhibits the warm phase PDF
depicted in Fig.\,\ref{fig:Bpdf_hwc}, apart from at high velocities where the
hot phase is more visible. The PDF is approximately Gaussian for the cold phase.
Based on this component separation, the sharp PDF profile of magnetic field in the hot phase, and less so for the warm, is likely connected with the
large-scale compressive forcing in the warm and hot phases, and subsequent
turbulent cascade.
The cold phase contains a magnetic field with a distribution that is between an
exponential and a Gaussian, resembling that of
\PXX, Fig.\,11, apart from having a weaker magnetic field strength.
It is possible that the higher densities in the cold phase may act as a 
sponge for these compressions and rarefactions, but we cannot assume that the
effects of the turbulent cascade driven by SN are absent even at the scale of molecular clouds.

Therefore, some caution should be attached to the velocity PDF for the cold phase
for at least two reasons.
In their Fig.\,15 (c) \citet[][]{gent2013I} show that the hot phase flows are
 mainly subsonic, the warm transonic and the cold supersonic. 
However, the cold clusters are typically entrained within the bulk
flows of their ambient warm gas.
If the bulk velocity of the ambient warm gas were subtracted, then the Mach
number of the cold phase would likely reduce, and the residual flow might
retain more of the PDF structure of the hot and warm phases.
Also, in this model the scale of the cold structures tend to be only a few 
grid spaces across, i.e., they are near the limit of the model resolution, so
much of the substructure of the magnetic and velocity fields in this phase are
truncated.
So although the MHD model here is truncated at 4\,pc, it is our contention 
that the physics that drive the structure of the magnetic field in the hot and
warm phases are still relevant to the flow driving the dynamics at smaller 
scales.
This would require comparison with higher resolution multi-phase turbulence simulations.
Only from new physical processes, such as self-gravity, would we expect to
introduce changes to the structure of the turbulence. 

%-----------------------------------------------------------------------------
%-----------------------------------------------------------------------------
%-----------------------------------------------------------------------------
%-----------------------------------------------------------------------------

%-----------------------------------------------------------------------------
\section{Shock and magnetic structure interpretation}\label{sec:analysis}
%-----------------------------------------------------------------------------

We now consider how $\Sd$-filament structures seen in the polarization angle dispersion 
measurements are related to physical properties of the ISM. These are 
difficult to measure directly through observations, but can be measured 
easily in the MHD models.
In the analysis that follows, we mostly refer to integration along the LOS with 
$\Rmax=1$\,kpc. 
This range, within the properties and horizontal extent
of the MHD model, is sufficient to adequately capture the key features present in the 
\PIX\ observations.
For more demanding analysis it would be recommended to integrate
$\Rmax\simeq2L_x$.

%-----------------------------------------------------------------------------
   \begin{figure}
   \centering
      \begin{subfigure}[t]{0.4\textwidth}
         \includegraphics[width = \hsize]{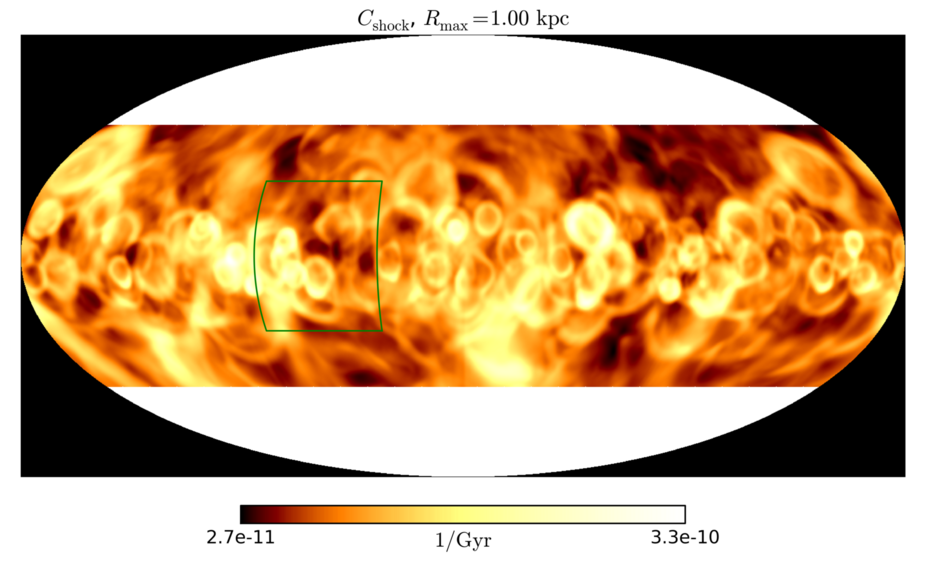}
      \end{subfigure}
      \begin{subfigure}[t]{0.4\textwidth}
         \includegraphics[width = \hsize]{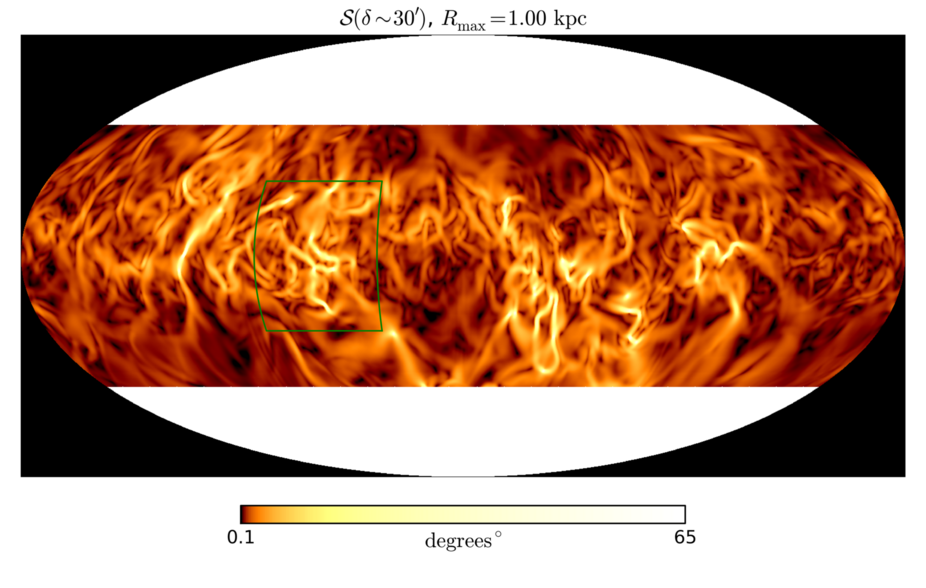}
      \end{subfigure}
      \begin{subfigure}[b]{0.4\textwidth}
         \includegraphics[width = \hsize]{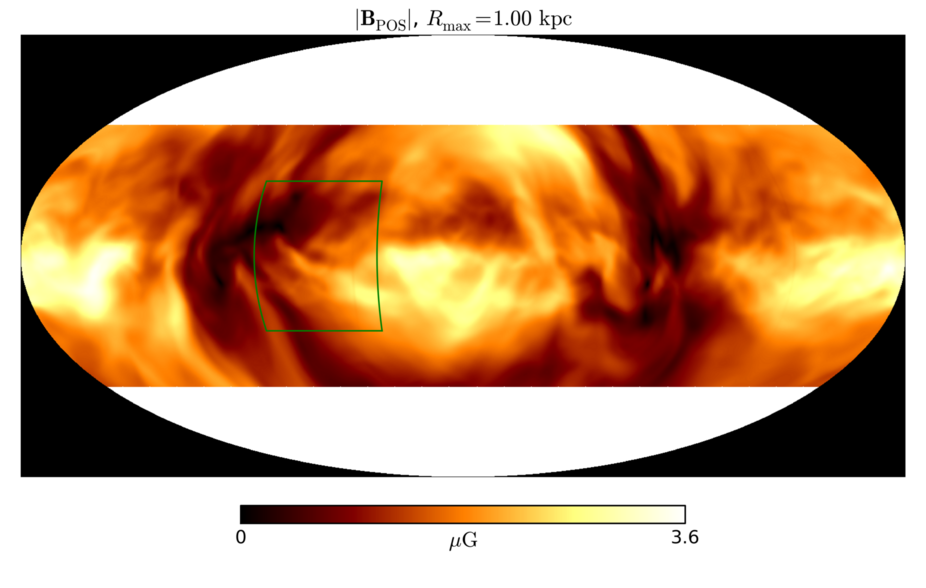}
      \end{subfigure}
      \caption{{\it Top:} The distribution of the average values of $C_{\rm shock}$ in the LOS. 
               {\it Middle:} Map of polarization angle dispersion $\Sd$.
               {\it Bottom:} Projected average magnetic field strength in POS.
               The green rectangles refer to the area featured in
               Fig.\,\ref{fig:local}.
               $\Rmax=1$\,kpc.
               }
         \label{fig:shock_compar}
   \end{figure}
%-----------------------------------------------------------------------------

%-----------------------------------------------------------------------------
\subsection{$\Sd$-filaments compared with shocks}
%-----------------------------------------------------------------------------

%-----------------------------------------------------------------------------
   \begin{figure*}
   \centering
      \begin{subfigure}[t]{0.4\textwidth}
         \includegraphics[width = 0.9\hsize]{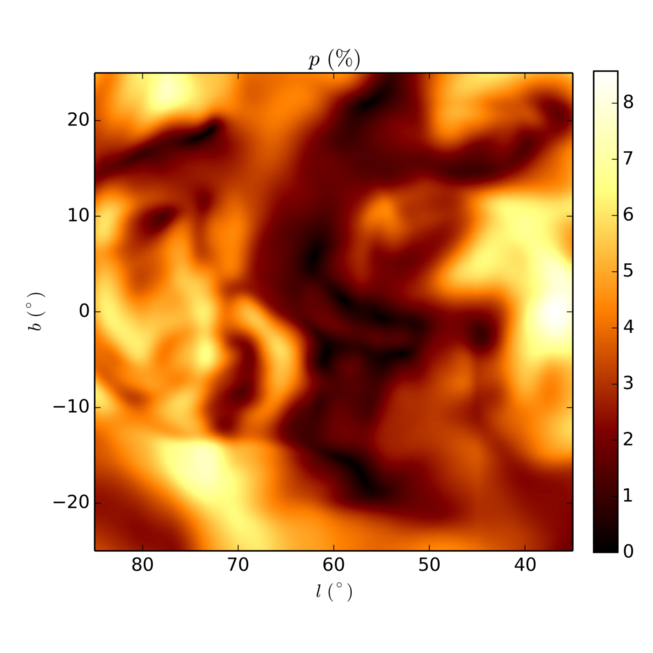}
      \end{subfigure}
      \begin{subfigure}[b]{0.4\textwidth}
         \includegraphics[width = 0.9\hsize]{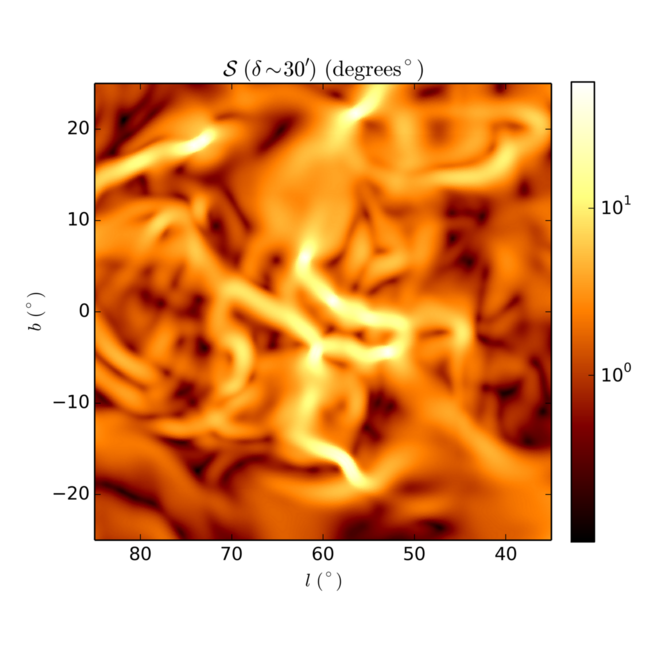}
      \end{subfigure}
      \begin{subfigure}[t]{0.4\textwidth}
         \includegraphics[width = 0.9\hsize]{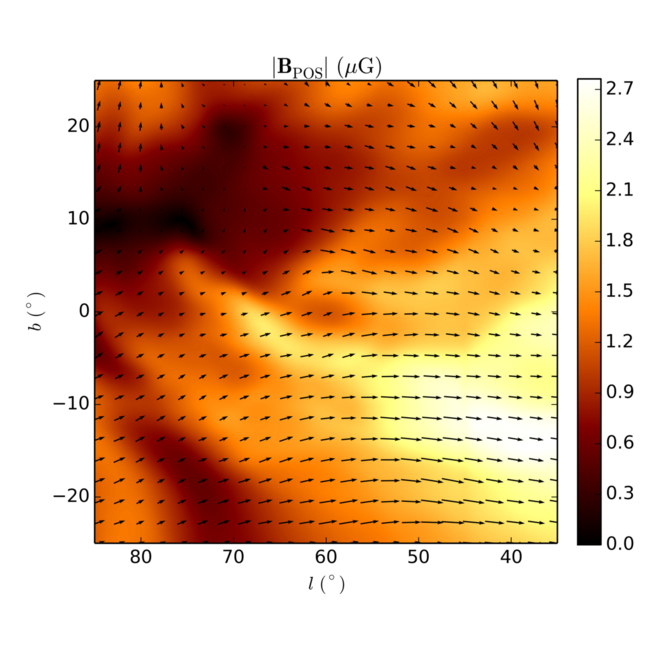}
      \end{subfigure}
      \begin{subfigure}[b]{0.4\textwidth}
         \includegraphics[width = 0.9\hsize]{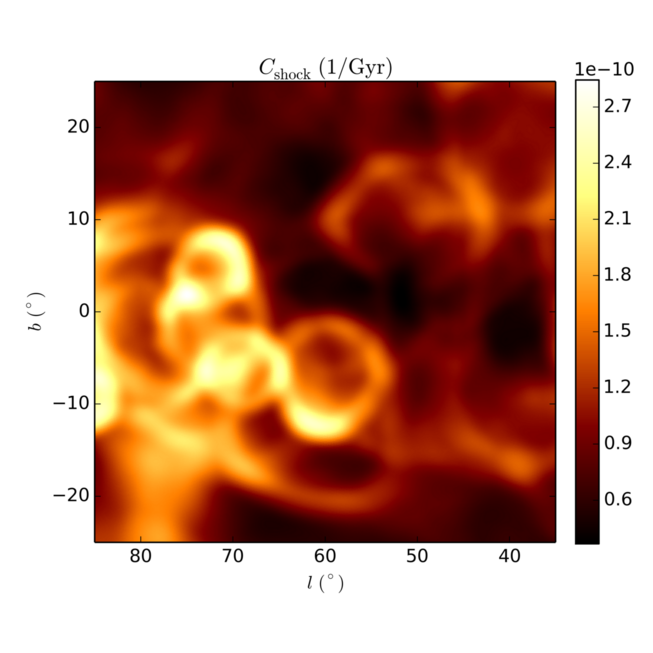}
      \end{subfigure}
      \caption{A local comparison of $p$, $\Sd$, $|\mathbf{B}_\mathrm{POS}|$, and shocks within the area marked in
               the Figs. \ref{fig:pol_pmap_flux} and \ref{fig:shock_compar}.
               }
         \label{fig:local}
   \end{figure*}
%-----------------------------------------------------------------------------

%-----------------------------------------------------------------------------
   \begin{figure*}
   \centering
      \begin{subfigure}[t]{0.35\textwidth}
         \includegraphics[width = 0.9\hsize]{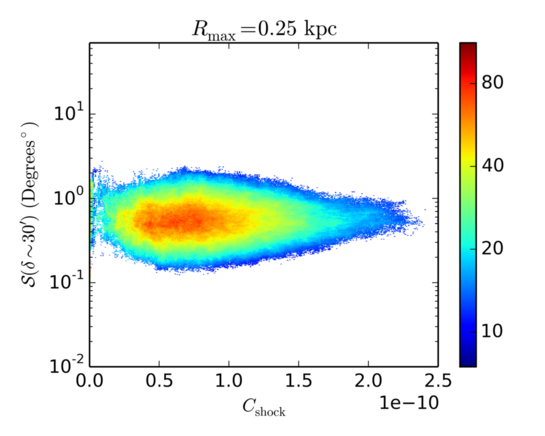}
      \end{subfigure}
      \begin{subfigure}[b]{0.35\textwidth}
         \includegraphics[width = 0.9\hsize]{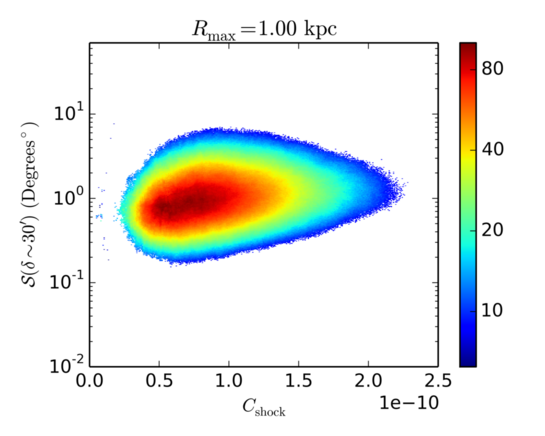}
      \end{subfigure}
      \begin{subfigure}[t]{0.35\textwidth}
         \includegraphics[width = 0.9\hsize]{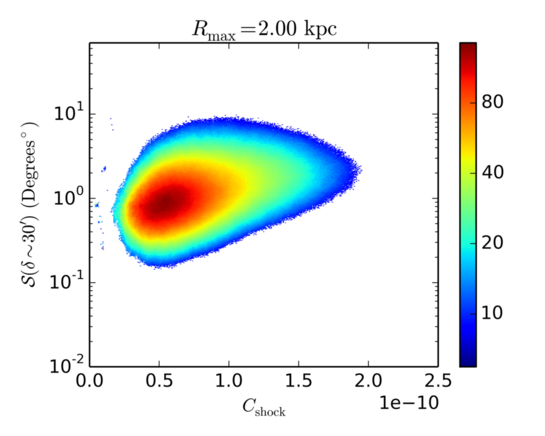}
      \end{subfigure}
      \begin{subfigure}[b]{0.35\textwidth}
         \includegraphics[width = 0.9\hsize]{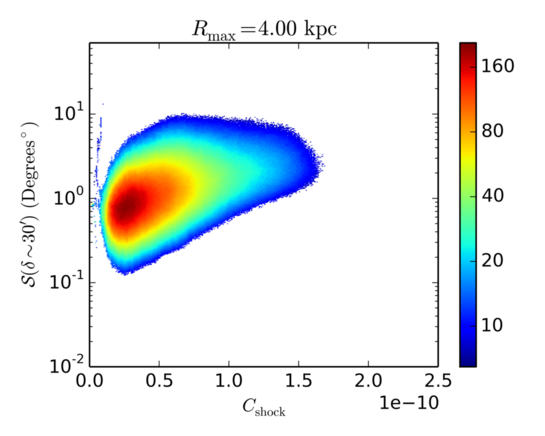}
      \end{subfigure}
      \caption{Joint 2D-histograms of polarization angle dispersion $\Sd$ and
                   average LOS shock compression $C_{\rm shock}$.} 
         \label{fig:hist_SH}
   \end{figure*}
%-----------------------------------------------------------------------------
 
Changes in the direction of polarization angle $\psi$ and therefore $\Sd$ are
related to changes in the magnetic field, and these are driven and generated by 
SN-driven turbulence.
Generally, $\Sd$ follows a lognormal distribution (see \PIX\, Fig.\,14). 
The lognormal nature of the $\Sd$ distribution in the observed and simulated
ISM appears consistent with the effect shocked
turbulence has on the statistics of the gas
density \citep[as noted in][]{V-S94,ES04,gent2013I}, which encourages us to look into the
connection between $\Sd$ distribution and shocks present in our simulation data.

To investigate the effect that the shocks have, we first compute a proxy of
their magnitude, $C_{\rm shock, local} = |[\nabla \cdot \mathbf{u}]_-|$, where only
the negative divergence contributes. 
This corresponds to regions where the flow is \emph{con}vergent,
where therefore the shocks created by SNe are compressing the surrounding ISM.
The values of $C_{\rm shock, local}$ are calculated within the numerical grid of 
the MHD model. All $C_{\rm shock}$ maps show 
$C_{\rm shock, local}$ averaged over the LOS up to the defined $\Rmax$, or 
\begin{equation}
C_{\rm shock} = \langle C_{\rm shock, local} \rangle_\mathrm{LOS}\Big|_0^{\Rmax}.
\end{equation}
In Fig.\,\ref{fig:shock_compar}, maps are displayed for 
average $C_{\rm shock}$, 
$\Sd$, 
and the POS magnetic field, $|\mathbf{B}_\mathrm{POS}|$, averaged over the LOS up to $\Rmax=1$\,kpc.
A zoom-in area is marked on each map, for which the local maps are displayed
in Fig.\,\ref{fig:local}. 
The joint 2D-histograms of $\Sd$ and $C_{\rm shock}$ are presented in the
Fig.\,\ref{fig:hist_SH}.

Apart from the energy input to the turbulence being stronger in the midplane
due to the general distribution of SNe, there is no visible correlation
between the SN shocks and the average POS magnetic field nor polarization
angle dispersion. 
The 2d-histograms presented in the Fig.\,\ref{fig:hist_SH} do not show any
clear dependence between $\Sd$ and $C_{\rm shock}$, apart from the effect of 
midplane-weighted SN-distribution being more pronounced.
However, there is a large-scale pattern in the
magnetic field, which is discussed in the Sect.\,\ref{sec:largesdcale}. 

Examining the zoomed-in area displayed in Fig.\,\ref{fig:local}, 
the filamentary structures in $\Sd$ 
(top right panel)
overlap very sharply with the areas of low polarization fraction
(top left panel).
There is likely a connected phenomenon, which links these effects.
Directly relating this to specific physical features in the model is a 
challenge, because locally the alignment of polarization is a combination of
effects layered on the top of each other.
One approach to understanding this is to perform a series of 
calculations over a range of discrete integration lengths $\Rmax$ and to analyse
in detail how the maps change as certain features of the model are included or
excluded.
It is also possible to explore the relationship between magnetic fields and
dispersion over large scales.

We stress that in advocating such an approach, we do not aim to explain
directly particular observational features from Planck, but to inform how the
inclusion or absence of physical phenomena along the LOS, which we can identify
precisely in the MHD models, might be expected to contribute to the simulated 
observations. 
Such results can also be used to deepen the physical interpretation of
the Planck findings.

When comparing the shock profile in Fig.\,\ref{fig:local}
(bottom right panel)
with the polarization angle dispersion 
(top right panel)
the strongest filamentary structures correspond to locations where the shocks
are negligible.
In the upper half of Fig.\,\ref{fig:local}
(bottom left panel)
the strength of the POS magnetic field seems to correlate quite well with the
polarization fraction 
(top right panel),
however in the lower right quadrant of the map a strong field is
anti-correlated to $p$, so the relationship is not at all straightforward.
In principle, all of these relationships should be explored further by 
varying $\Rmax$ as described above, but it appears likely that we can exclude the 
$\Sd$-filaments being indicative of the shock structure of the ISM. 

%-----------------------------------------------------------------------------
\subsection{Dispersion and the large-scale magnetic field}\label{sec:largesdcale}
%-----------------------------------------------------------------------------

%-----------------------------------------------------------------------------
   \begin{figure*}
   \centering
      \begin{subfigure}[t]{0.35\textwidth}
         \includegraphics[trim=0.0cm 0.0cm 0.0cm 0.0cm,clip=true,width = \hsize]{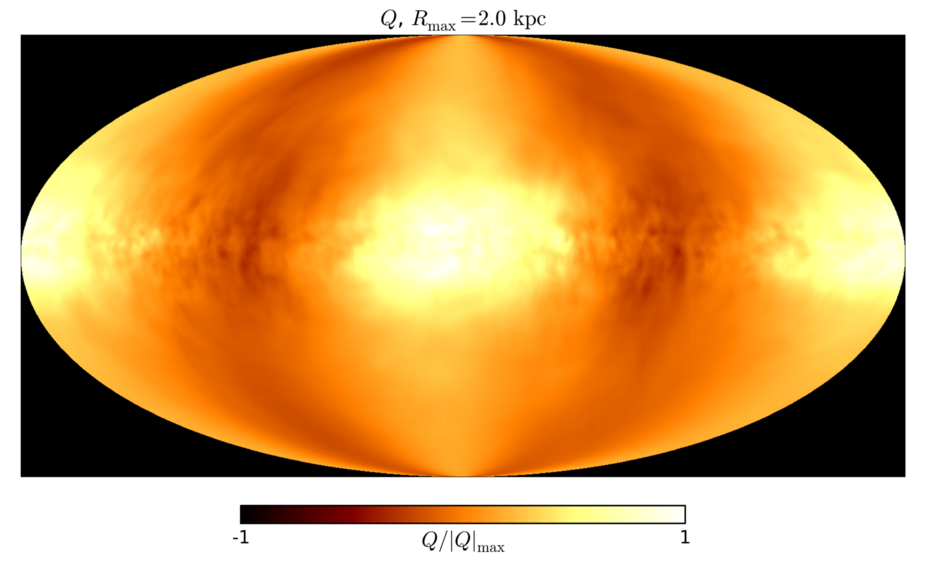}
      \end{subfigure}
      \begin{subfigure}[t]{0.35\textwidth}
         \includegraphics[trim=0.0cm 0.0cm 0.0cm 0.0cm,clip=true,width = \hsize]{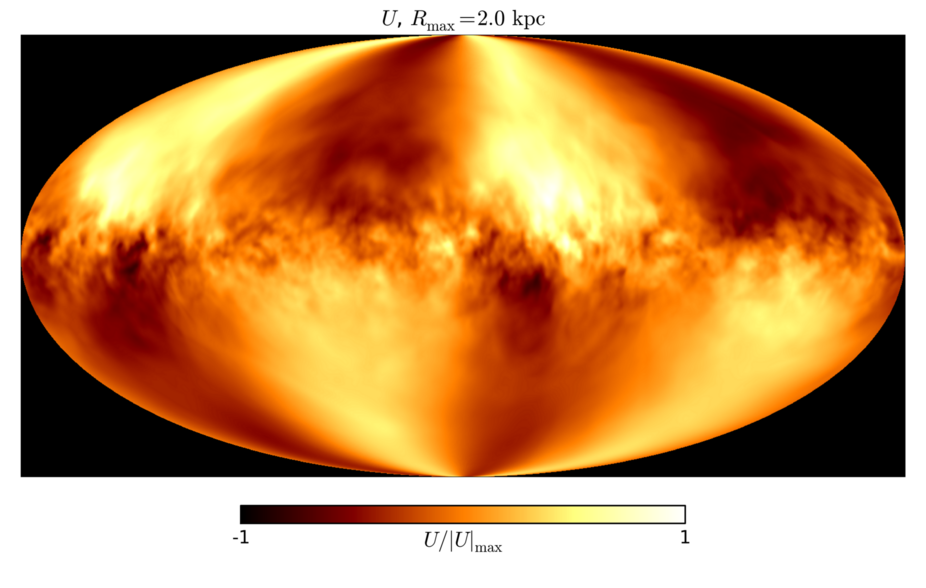}
      \end{subfigure}
      \begin{subfigure}[b]{0.35\textwidth}
         \includegraphics[trim=0.0cm 0.0cm 0.0cm 0.0cm,clip=true,width = \hsize]{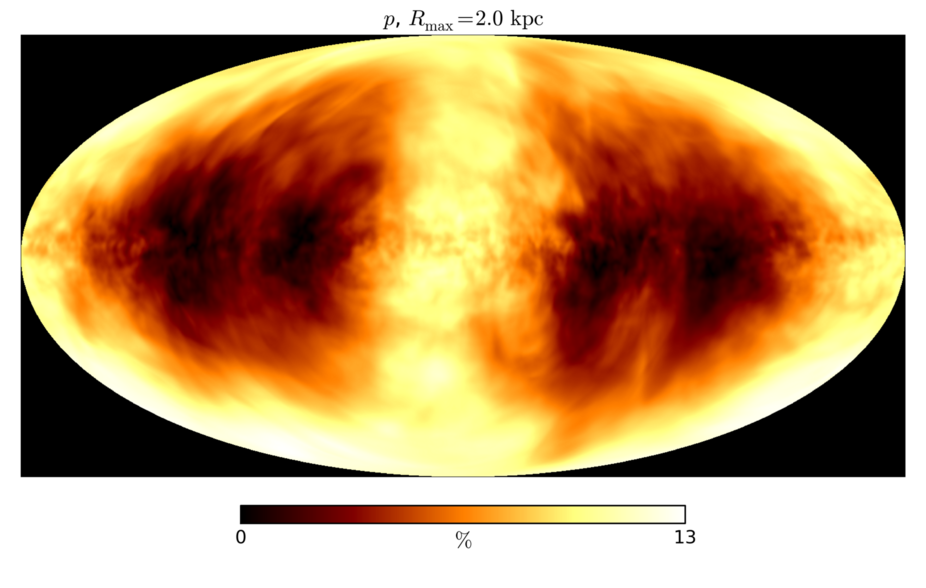}
      \end{subfigure}
      \begin{subfigure}[b]{0.35\textwidth}
         \includegraphics[trim=0.0cm 0.0cm 0.0cm 0.0cm,clip=true,width = \hsize]{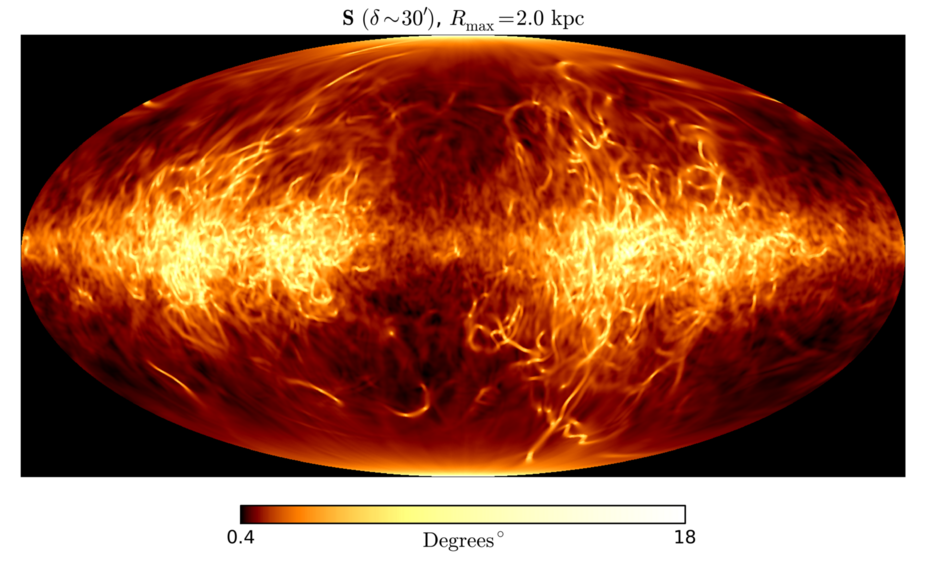}
      \end{subfigure}
      \caption{Averaged maps of $Q$, $U$, $p$, $\Sd$ over the all 12 snapshots.  
               }
         \label{fig:pol_avers}
   \end{figure*}
%-----------------------------------------------------------------------------

%-----------------------------------------------------------------------------
   \begin{figure*}
   \centering
      \begin{subfigure}[t]{0.35\textwidth}
         \includegraphics[width = \hsize]{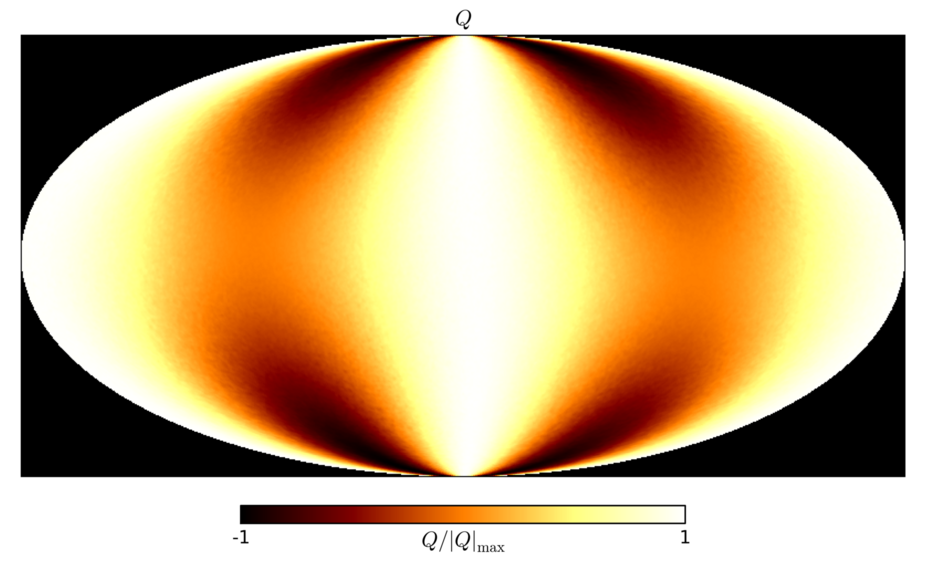}
      \end{subfigure}
      \begin{subfigure}[t]{0.35\textwidth}
         \includegraphics[width = \hsize]{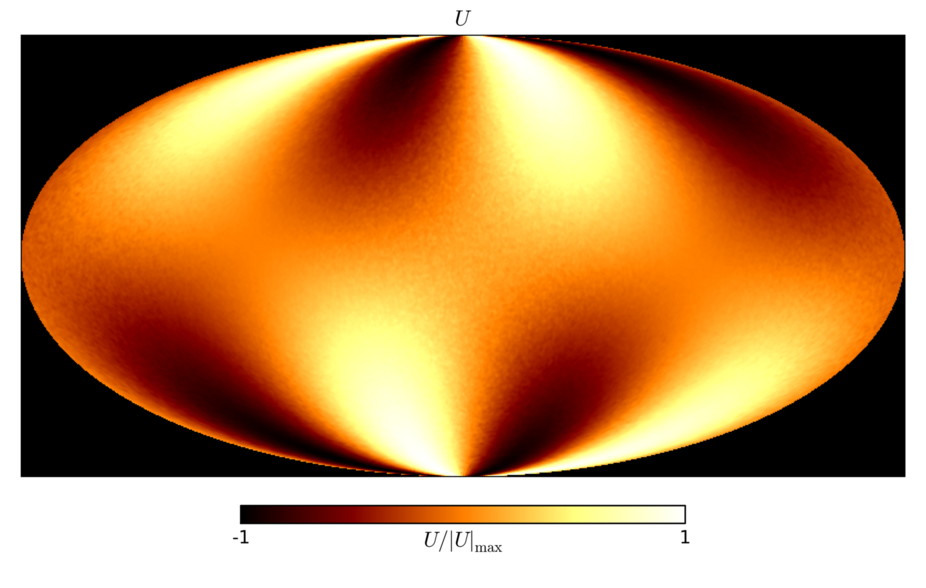}
      \end{subfigure}
      \begin{subfigure}[b]{0.35\textwidth}
         \includegraphics[width = \hsize]{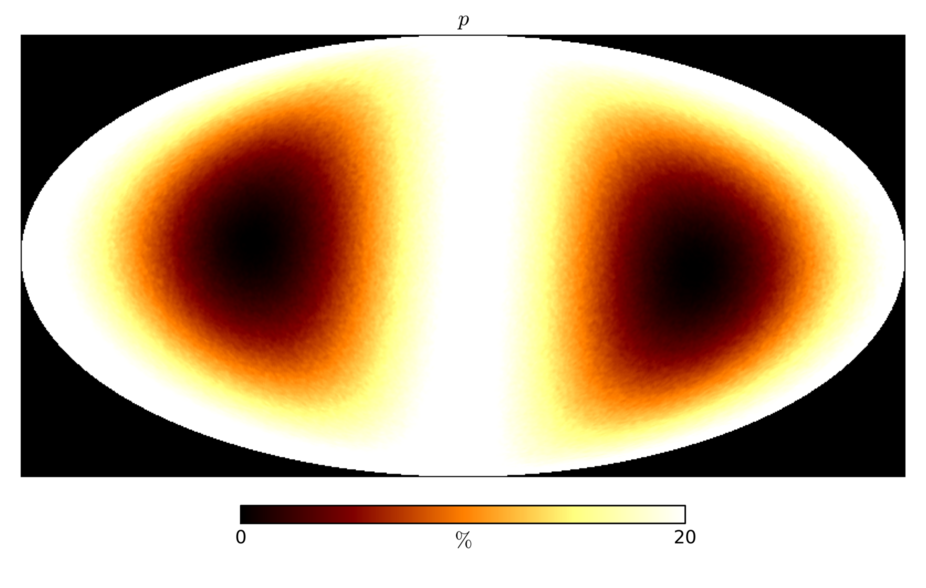}
      \end{subfigure}
      \begin{subfigure}[b]{0.35\textwidth}
         \includegraphics[width = \hsize]{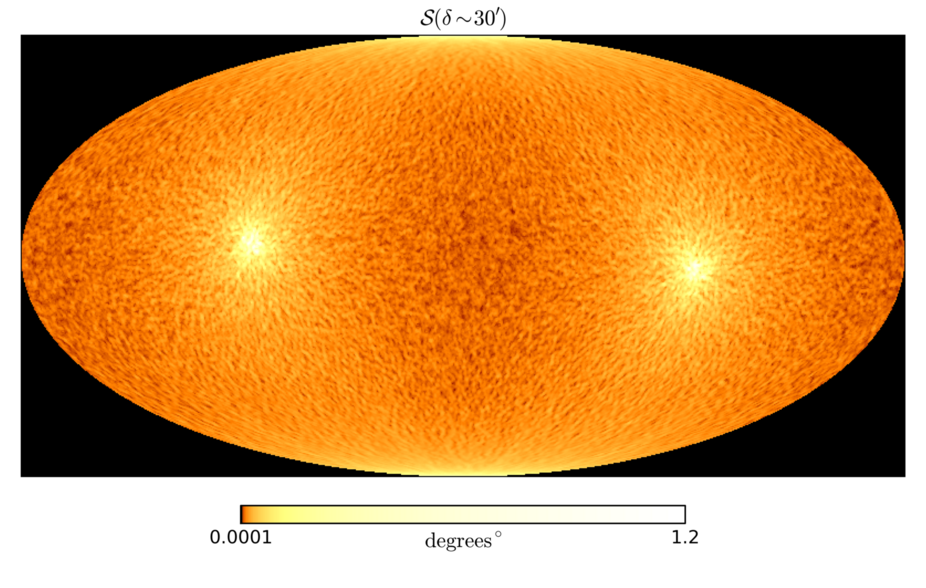}
      \end{subfigure}
      \caption{Distribution of polarization assuming 
	       a simple analytical model of the magnetic field
	       as described by the Eq.
               \ref{eq:B_meanfield}.
               }
         \label{fig:simple_MF}
   \end{figure*}
%-----------------------------------------------------------------------------

Already \citet{heiles1996}, for example, have used starlight polarization measurement to
estimate the direction and curvature of the galactic magnetic field. Their
results showed signs of spiral curvature, although the fluctuating component of
the field is also significant. More recent studies
\citep{Planck2016XLII,Planck2016XLIV} have used the new Planck results to estimate
the structural properties of the Galactic magnetic field
by fitting magnetic field models to observations.

In our synthetic observations, the distribution of polarization angle
dispersion exhibits a significant dependence on galactic latitude and longitude.
This becomes more apparent when taking averages from all simulated
observations with the same observer location, but for different snapshots, 
as is presented in Fig.\,\ref{fig:pol_avers}
(lower right panel).
The measurements of $Q$ reverse twice in one full rotation in latitude (upper
left panel),
while $U$ switches sign across the galactic plane and also exhibits the same
latitudinal sign reversals as $Q$ (upper right panel).
The polarization fractions are minimised where the brightest filamentary
structures are most pronounced.
The general nature of this pattern may be expected.
As outlined in Sect.\,\ref{sect:mhdsim}, the magnetic mean field is strongly
aligned in the direction of the differential rotation of the model.

The dependence of polarization properties on the orientation of the mean
magnetic field is a known relation. Already, in the context of turbulent molecular
clouds, \citet{ostriker2001}, \citet{SHMMNF13} and \PXX\ have shown that distribution
of fluctuations in polarization is connected with the direction of the magnetic mean
fields of molecular clouds. Such analysis has also been utilized with Planck
observation in relation to the molecular clouds \citep{Planck2016XXXII, Soler2016}.
This urges us to look into this phenomenon with our modelling results. However, as we
look into large-scale patterns, \citet{Planck2016XLII,Planck2016XLIV} provide
the most fruitful points of comparison.

The upper panels of Fig.\,\ref{fig:pol_avers} are remarkably consistent
with \citet[][Fig.\,13]{Planck2016XLII}, who examine a set of
galactic magnetic field models  without 
dynamo nor SN-driven turbulence to generate a realistic field, but including
the galactic centre and spiral arms, which they fit to observational data.
Their synthetic maps do not have the small-scale features associated with the 
turbulence, but on larger scales our maps have very similar structure, the
only characteristic difference being longitudinal variation caused by the included spiral arms.  
So, apparently, local structures in the ISM may be predominant in the 
observed data, although further investigation along the lines of
\citet{Planck2016XLII} and inclusion of spiral arms in a model of MHD turbulence
would need to be pursued.

%-----------------------------------------------------------------------------
   \begin{figure*}
   \centering
      \begin{subfigure}[t]{0.35\textwidth}
         \includegraphics[width = 0.9\hsize]{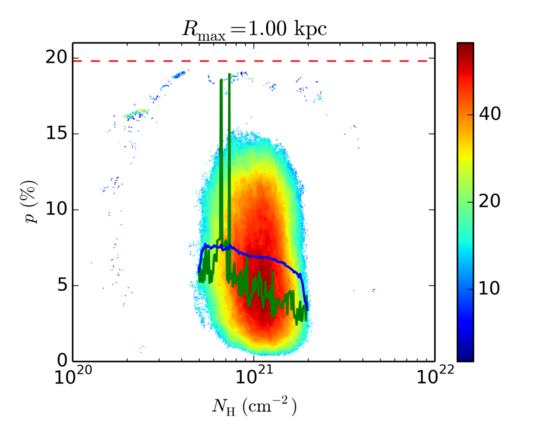}
      \end{subfigure}
      \begin{subfigure}[b]{0.35\textwidth}
         \includegraphics[width = 0.9\hsize]{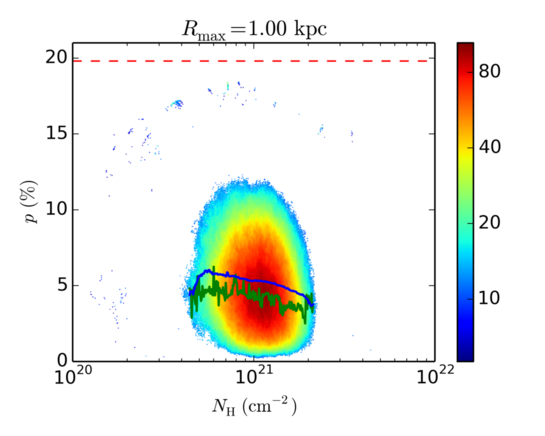}
      \end{subfigure}
      \begin{subfigure}[t]{0.35\textwidth}
         \includegraphics[width = 0.9\hsize]{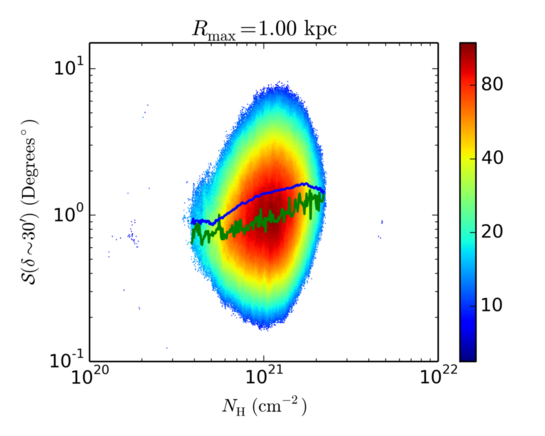}
      \end{subfigure}
      \begin{subfigure}[b]{0.35\textwidth}
         \includegraphics[width = 0.9\hsize]{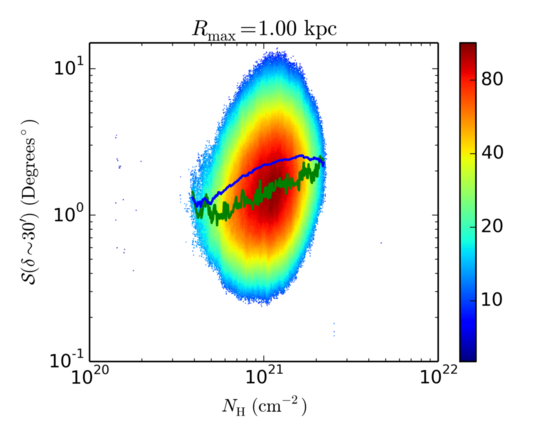}
      \end{subfigure}
      \begin{subfigure}[t]{0.35\textwidth}
         \includegraphics[width = 0.9\hsize]{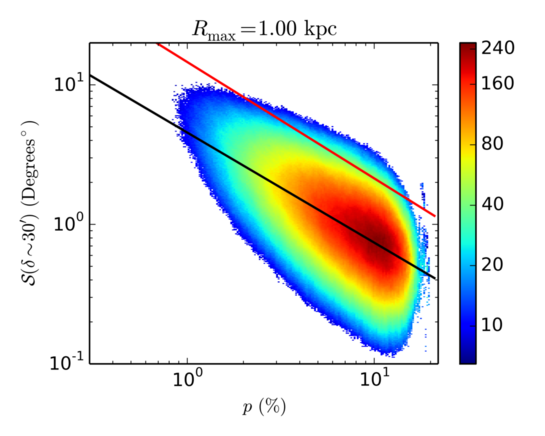}
      \end{subfigure}
      \begin{subfigure}[b]{0.35\textwidth}
         \includegraphics[width = 0.9\hsize]{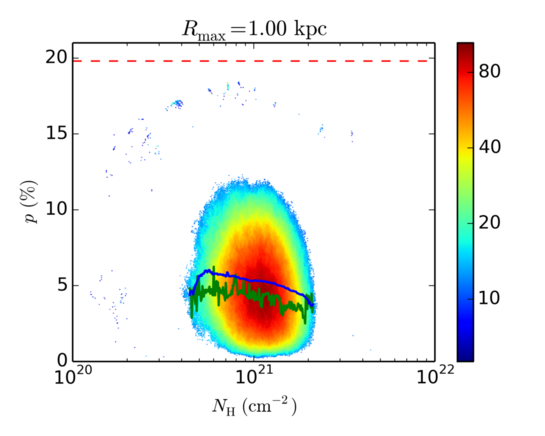}
      \end{subfigure}
      \caption{Joint 2D-histograms of ({\it Top}) polarization fraction and column
        density, ({\it Middle}) polarization angle dispersion and column
        density, and ({\it Bottom}) polarization angle dispersion and fraction.
        Red line depicts the fit to the observations as presented in
	\citet{Planck2015XIX}, $\log_{10} \Sd = \alpha \log_{10}p + \beta$. 
	Black lines are the best fits to our simulated data, to which $\alpha$
        and $\beta$ are given in the Table\,\ref{tbl:powerfit}. 
        Green lines follow the maximum counts in the 2D-histograms as a function
        of $\NH$, and blue lines present a weighted mean of the polarization
        fractions. 
        {\it Left:} Normal case. 
        {\it Right:} Doubled perturbations case.
        In all cases $\Rmax = 1 \, \mathrm{kpc}$.
        }
         \label{fig:hist_Np_flux}
         \label{fig:hist_pS_flux}
         \label{fig:hist_NS_flux}
   \end{figure*}
%-----------------------------------------------------------------------------

The large-scale variation follows from the presence of the mean field in
our MHD model. 
We can illustrate it with a simple analytical example starting from the first principles.
Let us assume a simple
uniform $y$-directional mean field with random fluctuations at the smallest
scales of the grid
\begin{equation}
\mathbf{B} = B_0 \hat{\bf y} + \Delta\mathbf{b}
\label{eq:B_meanfield}
\end{equation}
where $| B_0\hat{\bf y} | \gg | \Delta\mathbf{b} | $. 
This configuration
generates a large-scale 
structure of the polarization \citep[see Fig.\,\ref{fig:simple_MF}, and 
also][their Fig.\,4, top panels]{Planck2016XLIV}.
 In addition to this, the direction of the magnetic field
affects the sensitivity of the observed polarization to the small magnetic
fluctuations. 
If we apply Eq. \ref{eq:B_meanfield} to Eqs. \ref{eq:Bdir}, \ref{eq:psiangle}
and \ref{eq:cos2gamma} we notice
that near the HEALPix coordinates $\phi \approx \pm \pi/2$ and 
$\theta \approx \pi/2$ the influence
of the magnetic field approaches the values $\psi \approx \pi/2 + \Delta \psi$ 
and $\gamma \approx \pi/2 + \Delta \gamma$, where we have divided the contribution from the
mean and the fluctuating field. Therefore, when calculating the polarization
components, with Eqs. \ref{eq:Qpol} and \ref{eq:Upol}, 
we get,
\begin{equation}
Q \approx -I p_0 \cos \Delta \psi  \sin^2 \Delta \gamma
\end{equation}
\begin{equation}
U \approx -I p_0 \sin \Delta \psi  \sin^2 \Delta \gamma.
\end{equation}
This signifies that, when the LOS approaches the direction of the
consistent mean magnetic field, the POS field is highly sensitive to small,
local fluctuations caused by turbulence.
This, in turn,  will show up
as variations in polarization angle and therefore relatively high $\Sd$.
To summarize, when the strong mean field is perpendicular to the LOS,
its direction dominates the polarization angles, but when the field
is parallel to the LOS, the observed polarization angles are more
sensitive to the small fluctuations in the field. However, the polarization 
fraction $p$ is weak in the mean field aligned with the LOS, as the
small fluctuations themselves produce less strong $Q$ and $U$. 
Thus, we have a similar interpretation to
\PXX. 
In their study, $\Sd$ is strongest when the POV faces towards the mean field
direction, along with a weaker polarization fraction.
In contrast, they observe a higher polarization fraction and a coherent
polarization angle when the direction of the mean field follows the POS. 

%-----------------------------------------------------------------------------
   \begin{figure*}
   \centering
      \begin{subfigure}[t]{0.35\textwidth}
         \includegraphics[width = \hsize]{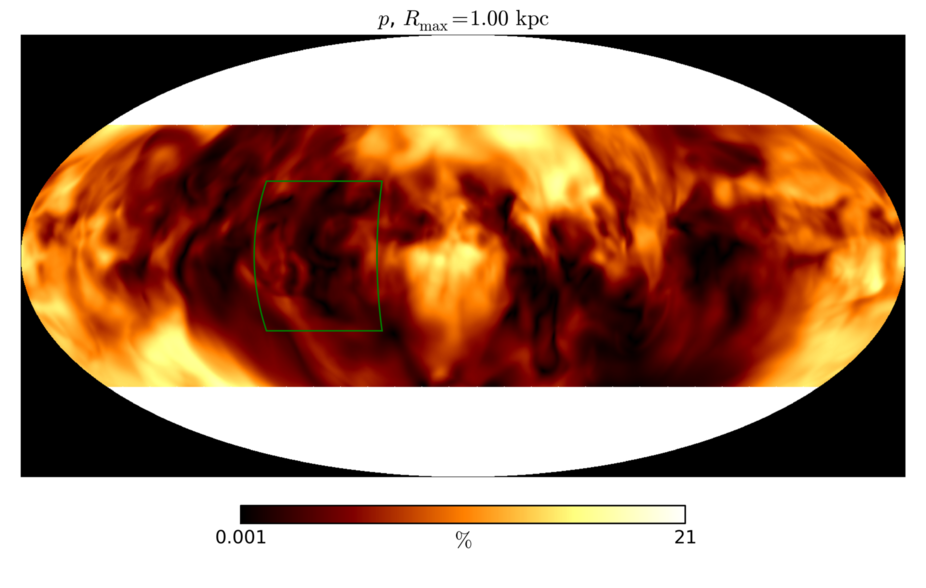}
      \end{subfigure}
      \begin{subfigure}[b]{0.35\textwidth}
         \includegraphics[width = \hsize]{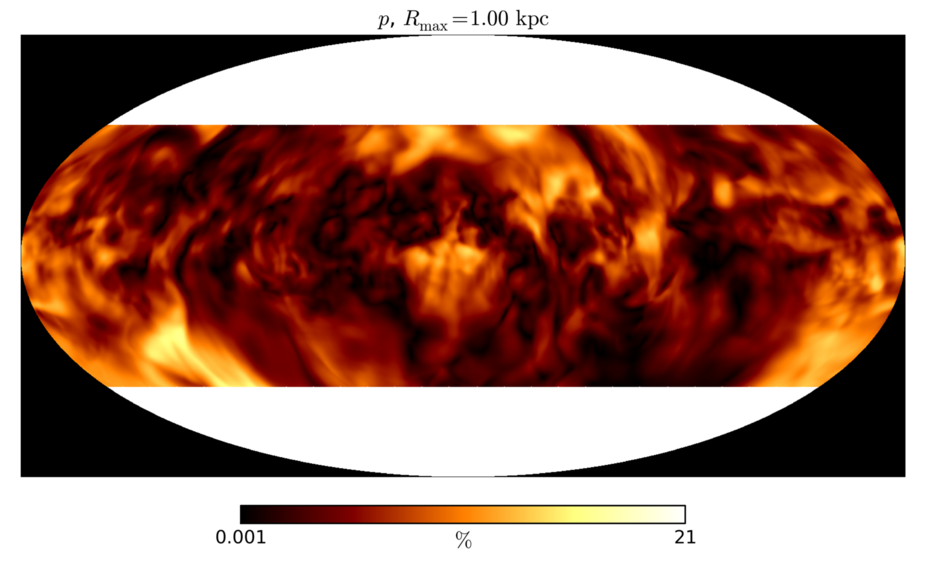}
      \end{subfigure}
      \begin{subfigure}[t]{0.35\textwidth}
         \includegraphics[width = \hsize]{pol_SS_1000_850_VAR11Final.png}
      \end{subfigure}
      \begin{subfigure}[b]{0.35\textwidth}
         \includegraphics[width = \hsize]{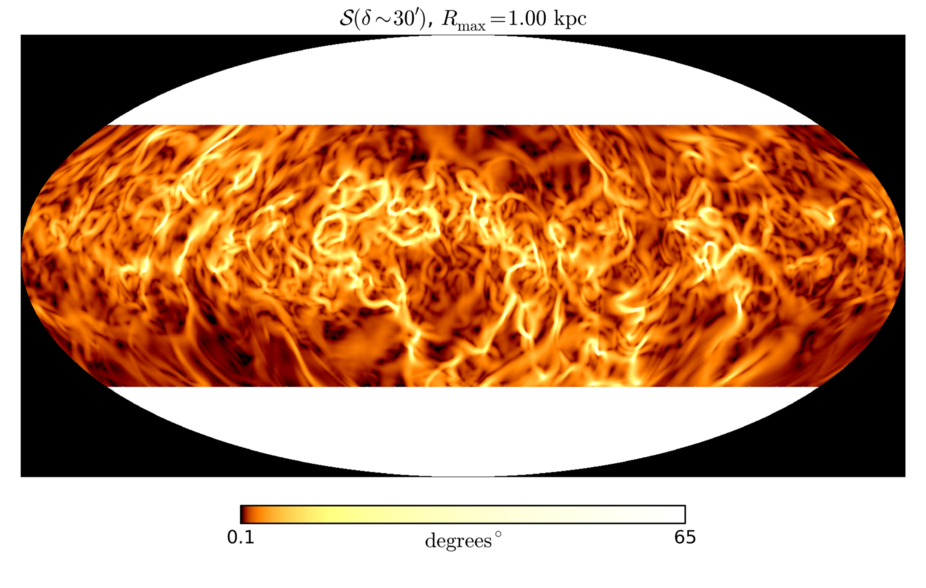}
      \end{subfigure}
      \caption{{\it Top:} Sample maps of polarization fraction $p$. 
               {\it Bottom:} Sample maps of polarization angle dispersion $\Sd$.
               {\it Left:} Normal case. 
               {\it Right:} Doubled perturbations case.   
               In all cases $\Rmax = 1 \, \mathrm{kpc}$,. 
               }
         \label{fig:pol_pmap_flux}
         \label{fig:pol_disp_flux}
   \end{figure*}
%-----------------------------------------------------------------------------

Note, that $B_0\hat{\bf y}$ here is distinct from $\bf{\bar{B}}$ as defined in 
Eq.\,\eqref{eq:meanB}. $\bf{\bar{B}}$ is defined by local averaging and includes
varying $x$ and $y$ components, although it is most strongly aligned along $y$. 
In the MHD model, the fluctuations $\bf{b}$ have the same order of magnitude
as $\bf{\bar{B}}$, which complicates visualising the large-scale mean field
even with the simulated observations. 
For observations, the structure of the mean field is even more 
opaque, as the mean field direction is subject to large-scale diversions 
when interacting with spiral arms and the central
bulge of the galaxy.
Therefore, using this interpretation to understand the observed $\Sd$ by
\PIX\ is not trivial. In the case of our MHD
simulation, the mean field is clearly stronger than the fluctuating field. In
the case of our Galaxy, the general structure of the large-scale field is more
complicated and the fluctuations are stronger \citep[e.g.][]{RK89,Haverkorn15}.
However, the effect of large-scale magnetic field on $\Sd$ could encourage
further study, as long as measurement error considerations will be taken into account
\citep{Montier2015I, AMRBLA16}.

%-----------------------------------------------------------------------------
\subsection{Effect of magnetic fluctuations}\label{sec:ef}
%-----------------------------------------------------------------------------

To assess the relevance of the mean and fluctuating field to the
synthetic observations, we explore the impact of increasing the amplitude of the
fluctuations relative to the mean field.
Using the decomposition of the field as illustrated in Fig.\,\ref{fig:3Dplots}, 
we double the amplitudes of the fluctuating component of the magnetic field and then sum it with
the mean field to obtain a physically generated field with stronger 
fluctuations. This means that as for the original dataset we get
$\sqrt{<b^2>}/\sqrt{<\bar{B}^2>} = 0.45$ and with the doubled amplitudes 
$\sqrt{<b^2>}/\sqrt{<\bar{B}^2>} = 0.90$.

Doubling the amplitudes of the fluctuating field somewhat reduces $p$ and increases $\Sd$,
as can be seen by comparing the left and right joint 2D-histograms in the
top two rows of Fig.\,\ref{fig:hist_Np_flux}.
The column density is not affected, understandable as the thermodynamic
properties of the model are unchanged.
The trends in the maximum count and weighted mean, traced on the 2D-histograms by the
green and blue lines, respectively, do not change, but they shift 
correspondingly with the general shift in the distribution. 
Comparing the joint 2D-histograms of $\Sd$ and $p$ in the last row of
Fig.\,\ref{fig:hist_Np_flux}, the log fit relating $\Sd$ to $p$ in the 
simulation shifts marginally closer to the \PIX\ fit, but the slope of the
line steepens.
In this sense the original MHD model still better explains the relationship between
$\Sd$ and $p$. 
The increases in $\Sd$ for the enhanced perturbations is associated mainly with 
low values of $p$, while $\Sd$ values associated with high $p\gtrsim10\%$ remain
less affected by the enhanced perturbations.

From the maps of $p$ in Fig.\,\ref{fig:pol_pmap_flux} 
(top row)
the polarization fraction is in general not only dampened, but also
exhibits increasingly fine structure.
This is also evident for $\Sd$ (bottom row), in which the filamentary structure
for the perturbed field with doubled amplitudes (right) is highly fractured, compared to the
map for the original field.
 
Increasing the relative strength of the fluctuations in the magnetic field
does not bring us visibly closer to the $\Sd$ values observed by \PIX.
Nor can longer integration distance address this, although this does
help with increasing column densities.
The distribution of polarization fraction in the MHD simulations is spread slightly higher than 
in \PIX, but doubling the strength of the random field component makes this a
very good fit to \PIX.
Part of the contribution to $\Sd$ comes from the mean field, which in the MHD
model is far from uniform. 
In the Galaxy, spiral arms and the central bulge will add to the variations
in the mean field.
Inclusion of such features in an MHD model would serve to enhance 
dispersion angles, but most likely the strongest factor is the limiting scale
of the magnetic field fluctuations and gas density concentrations.
These are truncated at the grid scale of 4\,pc. 
  
Given the long computation times as well as the large scales necessary to 
model a SN driven dynamo, increased resolution in the near future is likely to 
be modest.
Nevertheless, the trends and characteristics exhibited with these simulations 
have much in common with the \PIX\ results.
Exploration of these simulated observations has
 helped to reveal how 
different aspects of the magnetic field contribute to the observations.
A sweep of integration lengths $\Rmax$, comparing observations from 
many viewing angles, and comparing results across even a limited range of
MHD model resolution may offer further valuable insights.

%-------------------------------------------------------------------

%-----------------------------------------------------------------------------
\section{Discussion and conclusions}\label{sec:conclusions}
%-----------------------------------------------------------------------------

In this paper, using MHD models supplemented with radiative transfer
computations, we set out to study the effect on the
polarization of dust in the ISM of the following physical ingredients:
\begin{itemize}
\item SN regulated multiphase ISM, with hot component, and with longitudinal and
latitudinal anisotropy. 
\item The presence of ubiquitous shock fronts driven by SNe. 
\item The presence of self-consistently generated inhomogeneous and anisotropic
magnetic fields both by large- and small-scale dynamos. 
\end{itemize}
Previous investigations have been limited to the two-phase ISM, including 
only the cold molecular and diffuse warm gas components, \blue{either without
shocks or with} artificially
induced shock fronts, and with imposed magnetic field configurations.

We find a very good correspondence with the simulated $\Sd$ maps, exhibiting a strongly
filamentary structure, and the all-sky observations of 
\PIX\, implying
that our MHD models capture some essential features related to 
the formation of $\Sd$-filaments.
In accordance with the
observations, we find a good match to the anticorrelation between the polarization 
fraction $p$ and  the polarization angle dispersion $\Sd$.
The power law relation is quite accurately reflected in the simulation, 
although $\Sd$ differs by a factor 1/3 because it is sensitive to small-scale
fluctuations and the cold dense clouds, which our MHD model cannot 
sufficiently resolve. 

The mean magnetic field has both a systematic orientation in the direction of
the galactic shear and a non-uniform structure.
This significantly affects the observed polarization properties.
A strong plane-of-the-sky (POS) mean field is found to dominate over the
contribution of the small-scale component to the polarization angles in such a way
that the observed $\Sd$ is reduced.
Conversely, when the mean field is parallel to the line of sight (LOS),
the observed polarization angles are more sensitive to the small-scale
fluctuations.
Due to its varying orientation, the mean field also partially contributes to
the generation of $\Sd$-filaments.

In the light of these general findings, our key results can be listed as follows:
\begin{enumerate}
   \item We have demonstrated a means of probing the relationship between the
observations and the physical features along the LOS by varying the
integration length ($\Rmax$) in radiative transfer calculations.
   \item $p$ is correlated with the strength of the mean field in the  
POS and $\Sd$ is correlated positively with the fluctuating field $\bf b$
and inversely with $p$.
   We confirm the inverse correlation of the form $\log_{10} \Sd = \alpha \log_{10}p + \beta$, observed by \PIX. 
Our results support a view that a high polarization fraction indicates a strong 
POS coherent magnetic field, while a low polarization fraction is consistent with a strong
LOS mean field, supposing ${\bf{b}}$ to be approximately isotropic. It may be possible to apply this
to the measurements of $\Sd$ and $p$ to make inferences about the strength and 
orientation of the mean field.
    \item Filamentary structure of $\Sd$ becomes smaller scale, brighter and
more fragmentary with increasing $\Rmax$ as depolarization accumulates
along the LOS.
The general occurrence of brightest $\Sd$-filaments are well correlated with the large
scale shifts in POS orientation of the magnetic field.
\item
The $\Sd$-filaments do not correlate with the column
  density nor the location of SN shocks, but can be attributed to the distribution
  of the small-scale magnetic field.
  This is because the small-scale magnetic fields are the result of a small-scale dynamo
    also present in the MHD model, enabling field generation throughout the
    domain.
\item The fluctuations in the magnetic field are primarily driven by SN
	   turbulence and follow an exponential distribution in the hot and
		warm phase medium, while the cold phase medium follows a
		more Gaussian distribution. This indicates that the
		methods of model fitting assuming a
		Gaussian random magnetic field might not be the most sensible choice at
		least for the diffuse ISM, where the warm and hot components
		will play a part.
   \item Comparing joint 2D-histograms of $\Sd$ and column density $\NH$,
we probe the relationship
applying in real ISM between turbulent viscosity and $\Sd$.
We tentatively assert that turbulent viscosity in the fully ionized ISM reduces
slightly with increasing temperature as $T^{-\lambda}$, for some small positive
$\lambda$. Further MHD simulations with varying models of viscosity are required to test
this interpretation.
   \item We compare simulated polarization observations here and in
\citet{Planck2016XLII} with Planck observations.
It appears likely that the strong variations in \PIX\ observations are 
attributable to the physical structure of the ISM in the solar neighbourhood. 
Alternative models of the spiral structure and combining spiral arms with
SN turbulence MHD models are required to explore this further.
   \item Increasing  $\Rmax$ above 1\,kpc increases $\NH$
and reduces $p$ in line with observed distributions, suggesting that there is a
minimum value for $\Rmax$ which is needed for simulated observations.
However, to exclude artificial artefacts, we have to limit $\Rmax\lesssim2L_x$, where $L_x$ is
the model horizontal extent. 
\end{enumerate}

For future work, MHD models with increased resolution and/or horizontal extent
to probe the effects of longer integration ranges and higher densities and
fluctuations on $\Sd$ and $\NH$ would be helpful.
As these are essential for
many scientific priorities, these opportunities will undoubtedly be fulfilled.
Even with the current simulations the role of the fluctuating magnetic field
can be further investigated by retaining the existing MHD field and adding to
this an appropriate isotropic fluctuating field of the correct magnitude with
disturbances at various smaller scales.
Another way forward would be to use the saturated stages of the MHD models,
re-mesh them to include denser grids and thereby finer scales, and re-run 
only up to a new saturated stage, eliminating the long dynamo evolutionary
stage.
 
Conducting a series of experiments probing a range of $\Rmax$ to analyse how
physical features are being captured by the simulated observations in relation
to the POV would shed more light on how polarization features are connected
with the magnetic field structure.
Including spiral arms in MHD simulations and exploring how this impacts on the 
structure of the magnetic field and anisotropy in the synthetic polarization
observations could improve the interpretation of model results in terms of the
Milky Way galaxy.
Yet another possibility is to focus on zoomed-in features of the MHD models with
similarities to the observed features from Planck, and this way to explore what we
can learn about the 3D structure of the magnetic field at that location.

The often used Davis-Chandrasekhar-Fermi method \citep{Davis1951,CF53} allows for the
determination of the POS magnetic field strength if the dispersion of
polarization angle is known.
It is especially useful for estimating magnetic field strengths in regions,
such as molecular clouds, where the Zeeman effect is weak.
An important direction of future work is to test the predictions of this method
against self-consistently generated large- and small-scale fields,
ideally when the models can be refined to reach the limit where both
quantities resemble their observed counterparts.

As we have shown in this paper, the interplay of the large- and small-scale
magnetic fields can cause systematic effects in the polarization measures, that
may well be used to map the mean magnetic field of the Milky Way.
The ratio of $\Sd$ and $p$, reacting to the presence of different levels and
orientations of the magnetic field components, may be used as a tracer of the
orientation of the mean field, and on the other hand of the ratio of the
strengths of large- and small-scale magnetic fields.

\begin{acknowledgements}
MSV, FAG, and MJK acknowledge support of the Grand Challenge project SNDYN, 
CSC -- IT Center for Science Ltd. (Finland) and the Academy of Finland
Centre of Excellence ReSoLVE (project number 272157).
The simulations were performed using the supercomputers hosted by the CSC
-- IT Center for Science Ltd. in Espoo, Finland, which is administered by
the Finnish Ministry of Education.
MJ acknowledges the support of the Academy of Finland Grant No. 250741.
MSV thanks University of Helsinki and the Jenny and Antti Wihuri Foundation for financial support. 
Some of the results in this paper have been derived using the \verb+HEALPix+
package. 
This research has made use of NASA’s Astrophysics Data System.
\end{acknowledgements}

\bibliographystyle{aa} 
\bibliography{references} 

\begin{appendix}
\section{Maps of polarized emission}\label{sec:iqu}
\begin{figure}
   \centering
      \begin{subfigure}[t]{0.4\textwidth}
         \includegraphics[width = \hsize]{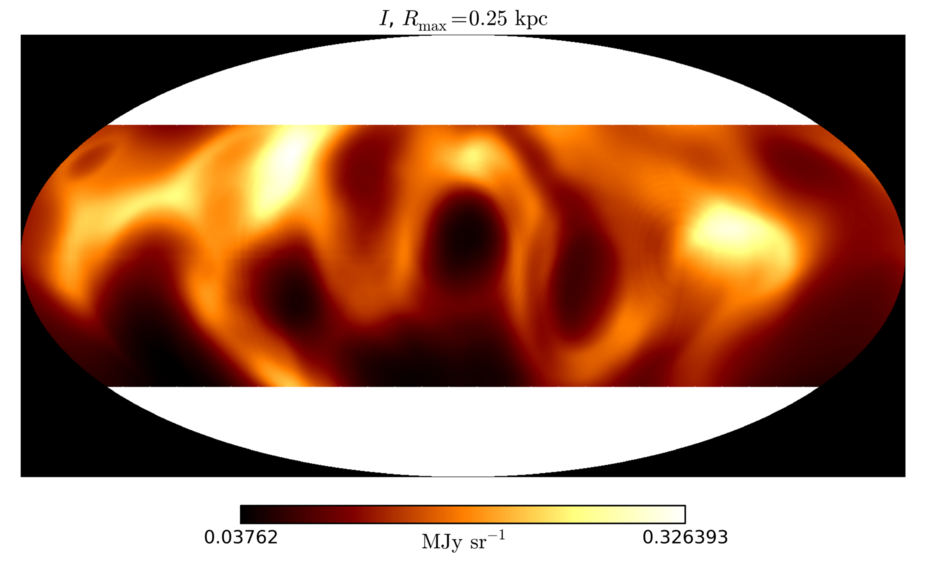}
      \end{subfigure}
      \begin{subfigure}[t]{0.4\textwidth}
         \includegraphics[width = \hsize]{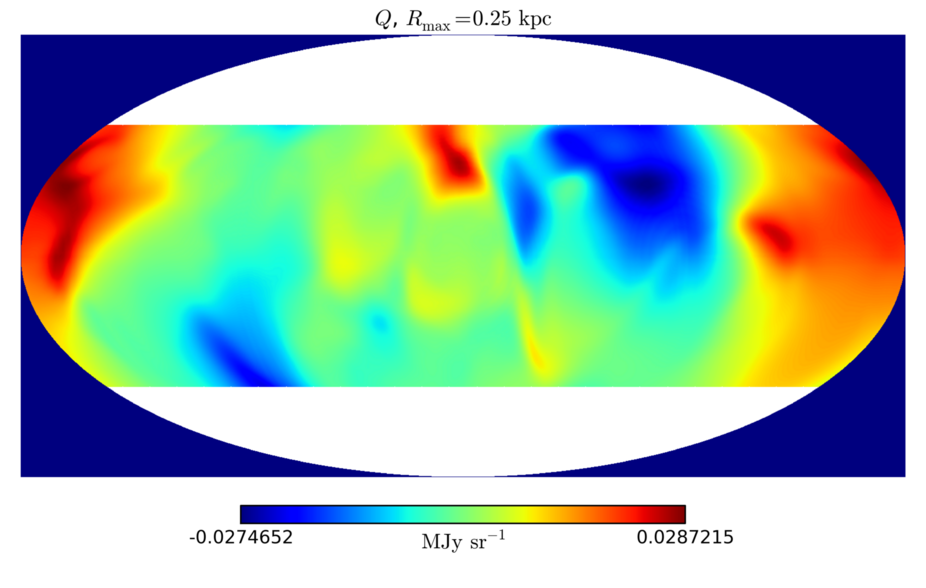}
      \end{subfigure}
      \begin{subfigure}[b]{0.4\textwidth}
         \includegraphics[width = \hsize]{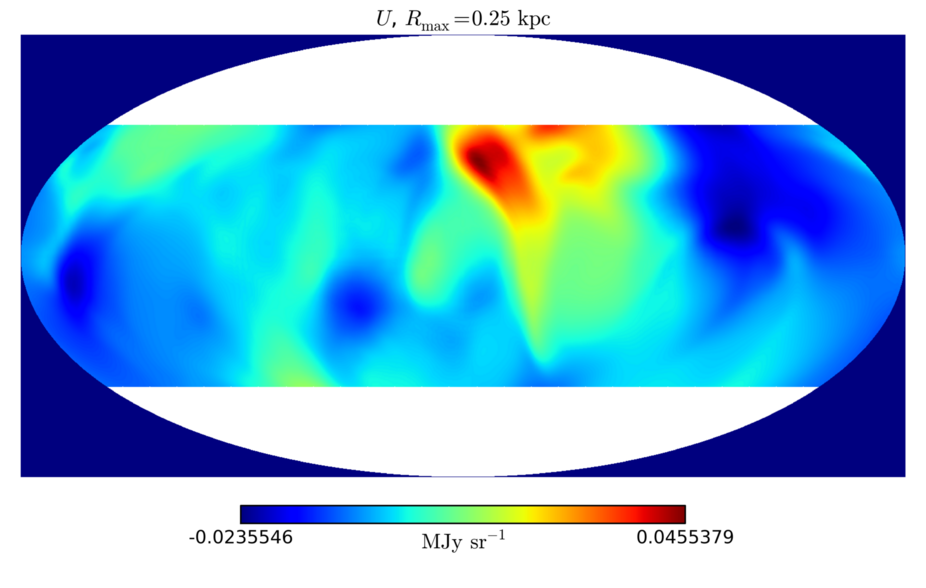}
      \end{subfigure}
      \caption{Stokes $I$, $Q$ and $U$ with $\Rmax=0.25$\,kpc.
               }
         \label{fig:R250}
   \end{figure}

\begin{figure}
   \centering
      \begin{subfigure}[t]{0.4\textwidth}
         \includegraphics[width = \hsize]{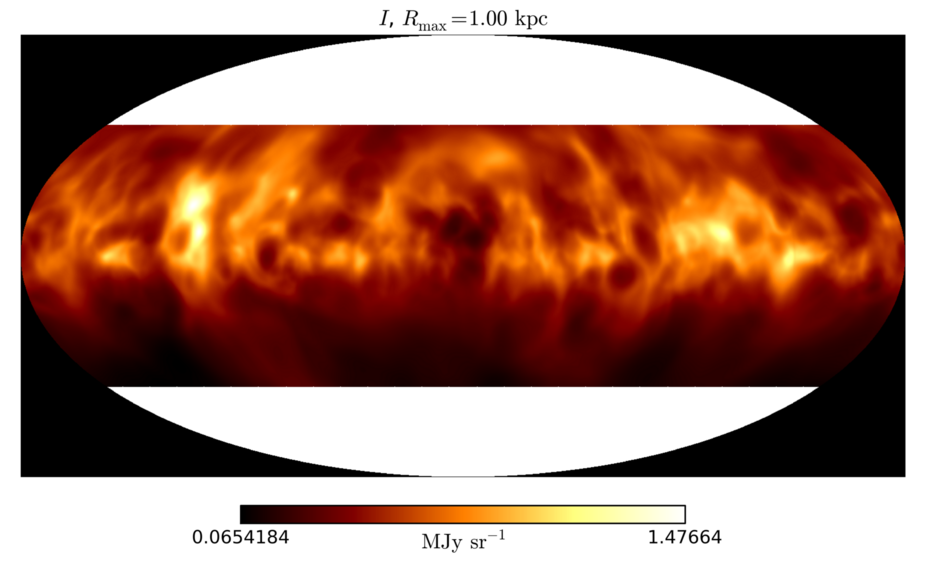}
      \end{subfigure}
      \begin{subfigure}[t]{0.4\textwidth}
         \includegraphics[width = \hsize]{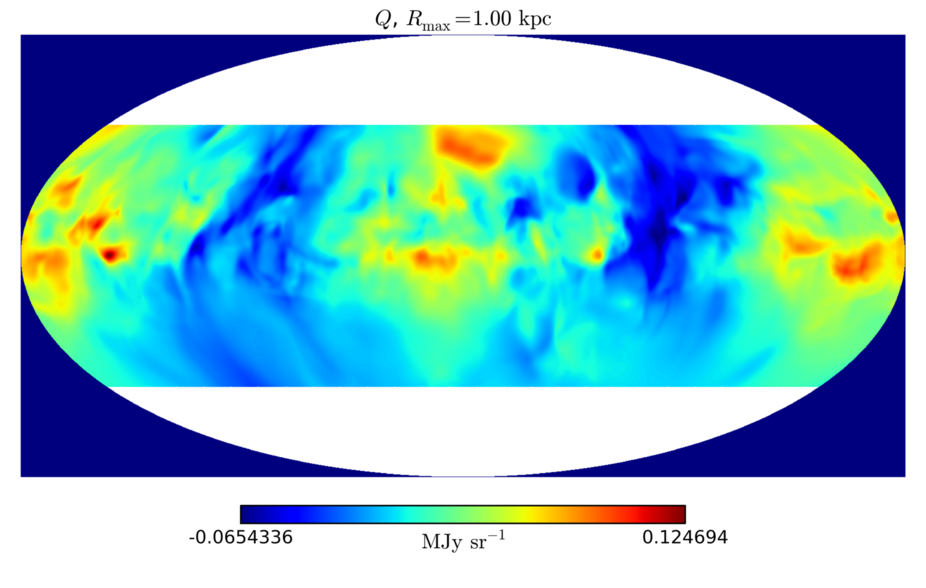}
      \end{subfigure}
      \begin{subfigure}[b]{0.4\textwidth}
         \includegraphics[width = \hsize]{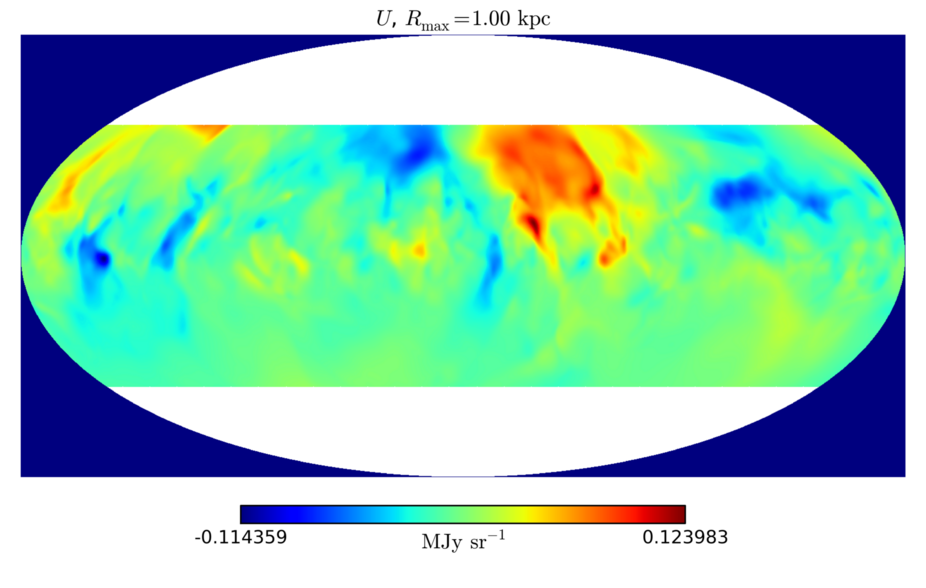}
      \end{subfigure}
      \caption{Stokes $I$, $Q$ and $U$ with $\Rmax=1$\,kpc.
               }
         \label{fig:R1000}
   \end{figure}

\begin{figure}
   \centering
      \begin{subfigure}[t]{0.4\textwidth}
         \includegraphics[width = \hsize]{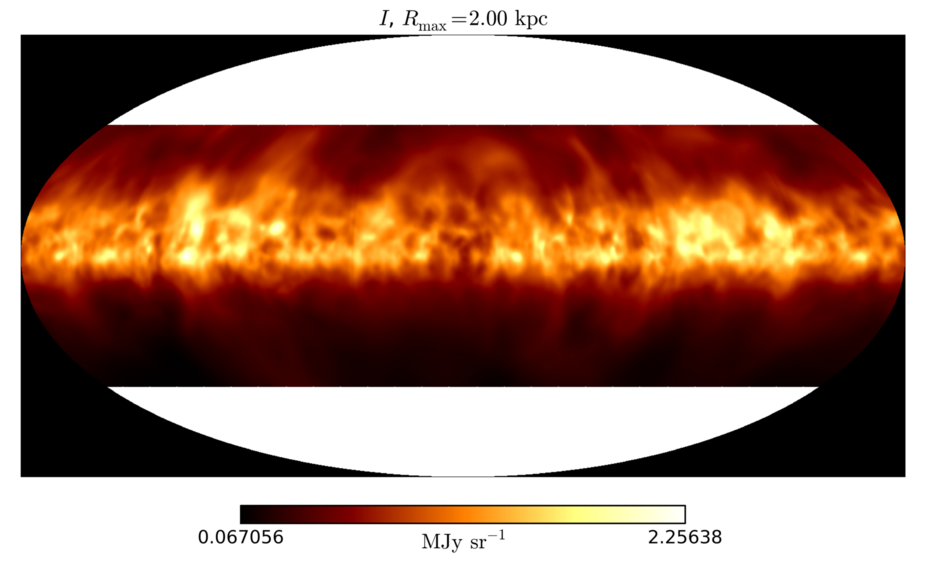}
      \end{subfigure}
      \begin{subfigure}[t]{0.4\textwidth}
         \includegraphics[width = \hsize]{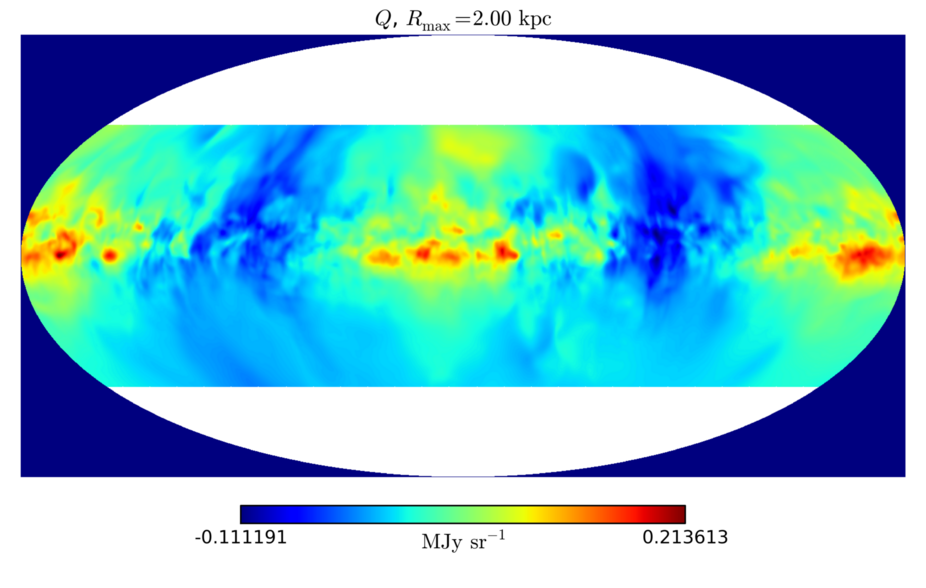}
      \end{subfigure}
      \begin{subfigure}[b]{0.4\textwidth}
         \includegraphics[width = \hsize]{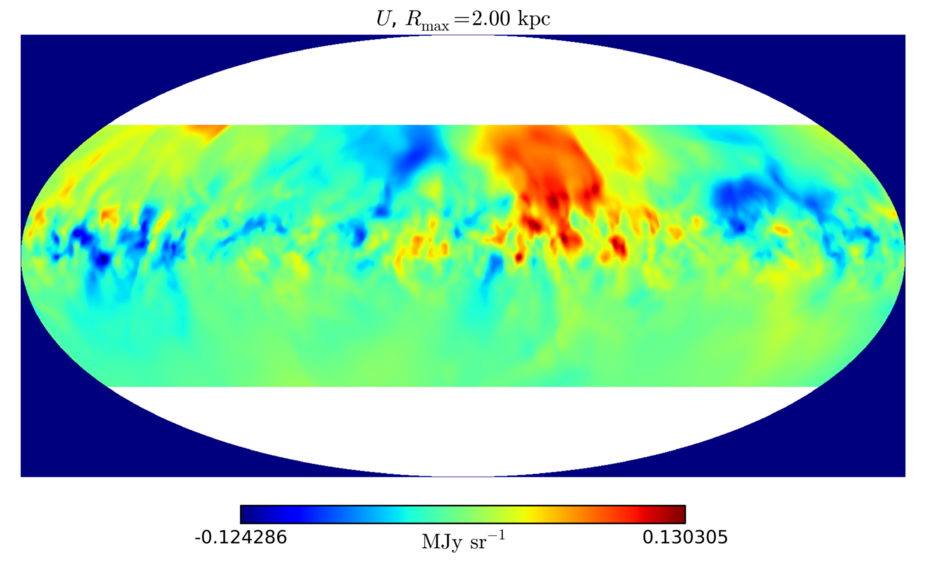}
      \end{subfigure}
      \caption{Stokes $I$, $Q$ and $U$ with $\Rmax=2$\,kpc.
               }
         \label{fig:R2000}
   \end{figure}

\begin{figure}
   \centering
      \begin{subfigure}[t]{0.4\textwidth}
         \includegraphics[width = \hsize]{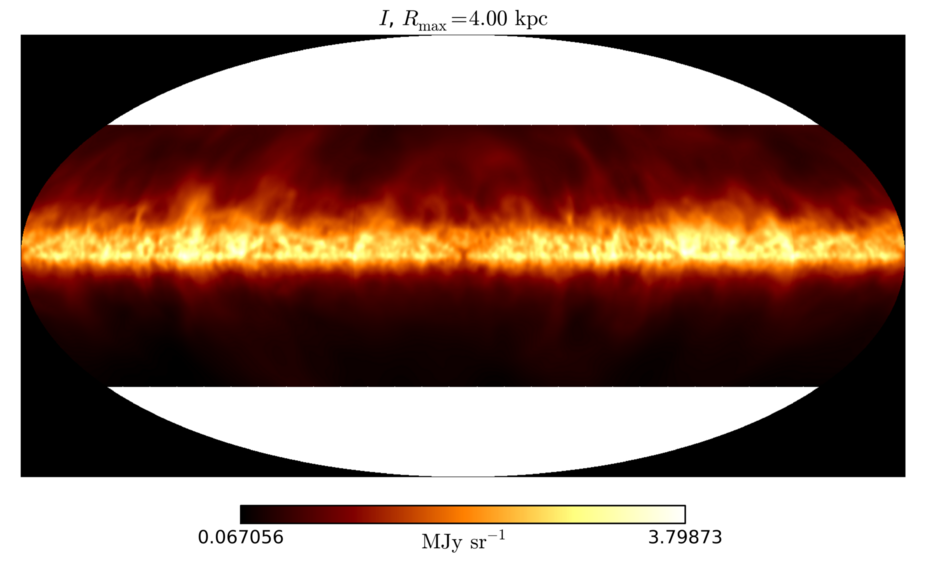}
      \end{subfigure}
      \begin{subfigure}[t]{0.4\textwidth}
         \includegraphics[width = \hsize]{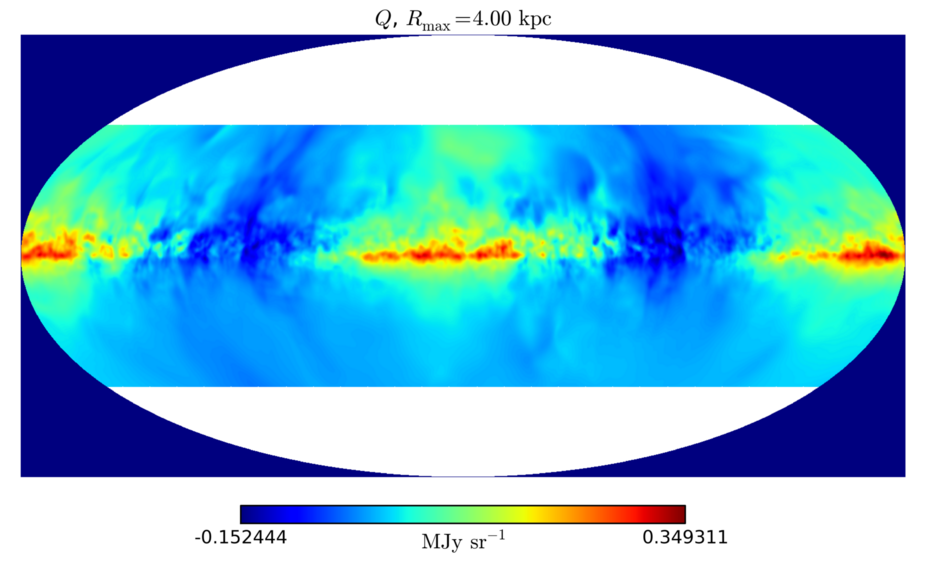}
      \end{subfigure}
      \begin{subfigure}[b]{0.4\textwidth}
         \includegraphics[width = \hsize]{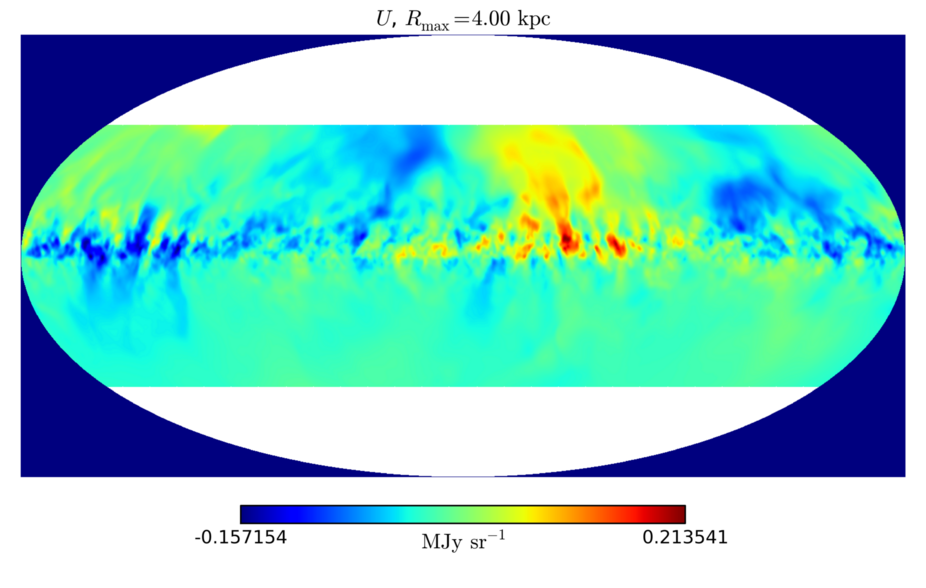}
      \end{subfigure}
      \caption{Stokes $I$, $Q$ and $U$ with $\Rmax=4$\,kpc.
               }
         \label{fig:R4000}
   \end{figure}

Here we feature maps of Stokes $I$, $Q$ and $U$ used for calculating $p$ and $\Sd$
shown by the displayed maps in Figs. \ref{fig:pol_pmap}, \ref{fig:pol_disp},
\ref{fig:shock_compar} and \ref{fig:pol_disp_flux}. 
See Figs. \ref{fig:R250}, \ref{fig:R1000}, \ref{fig:R2000} and \ref{fig:R4000} for
$\Rmax = 0.25$, $1.0$, $2.0$ and $4.0$ respectively.

\section{The MHD model}\label{app:model}
The focus of this paper is to explore the unique insights possible by using 
physically driven multiphase ISM and dynamo generated simulation magnetic field 
data to derive synthetic observations. 
These can then be compared with astronomical observations and those of
simulations based on imposed flows and magnetic field configurations.
We stress that the MHD simulations are not designed to model features in the 
ISM that have been observed or modelled by Planck or other observers, but 
independently to examine the multiphase structure and dynamo action of 
supernova driven turbulence in the ISM. 
These simulations have been described in detail elsewhere, but without breaking
the focus on the core purpose of this paper, here we shall clarify for
the reader more of the basic assumptions and techniques relating to the MHD
models.

In this particular model  SNe are modelled by injecting thermal and kinetic
 energy of 10$^{51}$\,erg into a spherical region, from which the pressure
gradient and momentum drive shock fronts and heating into the ambient ISM.
The thermal component has been found to be critical for efficient generation of
vorticity through baroclinicity \citep{KGVS17}, 
and this is beneficial for dynamo action,
unlike models applying momentum forcing, which are dominated by potential 
flows \citep{IH17}.
The timing and location are motivated by observational estimates of SN 
distribution in the solar neighbourhood, and placement is not dependent on the
local ISM density or temperature, avoiding excessive damping in dense material 
or the accumulation of immovable overdense regions.
The environment applies an initial vertical density profile based on 
estimates summarized in \cite{Ferriere01}, subject to gravitational 
acceleration following \cite{Kuijken89} and differentially rotating with
angular speed and rotation consistent with the solar neighbourhood. 
The resulting turbulence, magnetic field structure, and the multiphase
structure of the ISM are thus determined by the solution of the physical
equations and random SN events, and not imposed.
The system of equations being solved include continuity, momentum, energy
and induction treating the ISM as a monatomic ideal gas{\small{
\begin{eqnarray*}
  \label{eq:mass}
  \frac{D\rho}{Dt} & =&- {\rho \vec\nabla \cdot \vec{u}} + \dot{\rho}_{\rm SN},
\end{eqnarray*}
%----------------------------------------------------------------------------
\begin{eqnarray*}
  \label{eq:mom}
  \frac{D \vec{u}}{Dt}
  & = & -c_{s}^{2} \vec{\nabla} \left(\frac{s}{c_{p}} + \textrm{ln} \rho\right) +
  {\vec g} -S u_x\bm{\hat{y}} -2 \vec\Omega\times \vec{u} + 
  \frac{\vec{j}\times\vec{B}}{\rho}\\
  \nonumber
  &+ & \nu\vec \nabla^{2} \vec{u} + \frac{\nu}{3}\vec\nabla \vec\nabla \cdot
  \vec{u} + 2 \bm{\mathsf{S}} \cdot\left( \nu\vec{\nabla} \textrm{ln} \rho+\vec\nabla\nu\right) +\zeta_{\nu}
  \left(\vec{\nabla} \vec{\nabla}\cdot \vec{u} \right),
\end{eqnarray*}
%------------------------------------------------------------------------
  \begin{eqnarray*}
    \label{eq:ent}
    \rho T\frac{D s}{Dt}& =
    &\dot\sigma_{\rm{SN}}+\rho\Gamma-\rho^2\Lambda+c_p\vec{\nabla} \cdot \chi\rho \vec\nabla T 
    +2 \rho \nu \bm{\mathsf{S}}^{2}\\
      & +& 
  \rho T\vec{\nabla}\zeta_{\chi}\cdot \vec{\nabla}s + \mu_0\eta|\vec{j}|^2,
  \end{eqnarray*}
%------------------------------------------------------------------------
  \begin{eqnarray*}
    \label{eq:ind}
    \frac{\partial \vec{A}}{\partial t}&=&
    \vec{u}\times\vec{B} - S A_y\vec{\hat{x}} - S x \frac{\partial \vec{A}}{\partial y} \, \\ \nonumber
     &+&(\eta+\zeta_\eta)\vec{\nabla}^2\vec{A}
      +(\vec{\nabla}\cdot\vec{A})\, ({{\vec{\nabla}\eta+}}\vec{\nabla}\zeta_\eta),
  \end{eqnarray*}
}}with the symbols having their usual meaning. For a full explanation refer to the earlier papers.

Modelling a galactic dynamo required an integration time in excess of 
1\,Gyr, with a time step below 100\,yr, and on occasions less than 1\,yr,
with our chosen resolution of 4\,pc along each edge. 
A lower resolution would permit faster progress with reduced resources, but
careful experiments with various resolutions established 4\,pc provided
the minimum resolution with our numerical scheme to faithfully evolve SN 
remnants with the relevant essential physics \citep[][Appendix A]{Gent2012}.
There is a seed magnetic field of a few nG, which is amplified by the 
SN driven turbulence to saturate with a mean field strength a few $\mu$G. 
Unlike most other similar models with imposed magnetic fields and unphysical
forcing, the field structure is derived from the physical processes modelled.
The motivation for the 1\,kpc horizontal extent was to ensure no artificial
self-feedback from SN superbubbles, which can expand as far as 500\,pc, and
to include all estimated dynamo modes.
Some similar models apply AMR with local resolution near 1\,pc, although 
elsewhere within the models the resolution can be an order of magnitude 
weaker, but they are not seeking dynamo solutions.
Improving the resolution four fold would multiply the size of the calculation
at each iteration by $4^3$ and increase the number of iterations per Myr
by at least $4^2$.
A detailed analysis of the multiphase structure and properties of the ISM 
obtained with this model is reported in \citet{gent2013I}, and these are very
consistent with results reported by other authors adopting higher resolution
and employing AMR \citep{JM06,AB07,Hill12}.
Other than obtaining higher maximal values for density and higher proportion of
cold gas due to the enhanced resolution, the phases obtained are still
similarly located and distributed, and exhibit the same temperature and
velocity characteristics.

The reason for this is that the difference between 1 and 4\,pc does not
permit the inclusion of qualitatively distinct physics. 
Even a ten-fold increase in maximal densities would not bring the model within
the threshold demanding self-gravity. 
The Jeans mass for gas at 100\,cm$^{-3}$ at 100\,K is $4.4\times10^{33}$\,kg. 
The total in a $(4{\rm pc})^3$ volume element is less than $10^{31}$\,kg, so
no regions within the model are close to being susceptible to gravitational
instability, and this would be the case even for models with ten times more 
dense regions. 
The typical cooling times in our MHD model are about 100\,yr, but
occasionally as low as 1\,yr, never sufficiently dense to introduce 
unresolved thermal instabilities, which lead to cloud collapse and these
processes are also absent for models with 1\,pc resolution.
Even where isolated regions within such models may approach the critical Jeans
mass, they are unlikely to have enough significance to the global dynamics
to justify the numerical overheads.

Increased resolution is, of course, desirable, but comes with a numerical 
cost. 
Higher resolution would increase the Reynolds numbers applicable in the model,
permitting smaller magnetic and velocity structures to be resolved and perhaps
more efficient dynamo action.
For further illustrations, including video representations, of the simulation 
data see \href{http://fagent.wikidot.com/astro}{http://fagent.wikidot.com/astro}.

\end{appendix}

\end{document}